\documentclass[structabstract]{aa} 
\usepackage{graphicx}
\usepackage{upgreek}
\usepackage{amsmath}
\usepackage{mhchem}
\usepackage{float}

\usepackage{natbib}
\usepackage{twoopt}
\usepackage{multirow}
\usepackage{caption}
\usepackage{siunitx}
\usepackage{gensymb}
\usepackage{threeparttablex}
\usepackage{subfig}

\usepackage{mathtools}
\usepackage{braket}

\usepackage{txfonts}
\usepackage{color}
\usepackage{hyperref}
\hypersetup{%
  colorlinks=true,
  linkcolor=blue,
  citecolor=blue
}

\def \bmath #1 {{\hbox{\boldmath{$#1$}\unboldmath}}}

\usepackage[dvipsnames]{xcolor}

\begin{document}

\title{Formation of adducts of C$_6$H with Na$^+$, Mg$^+$ and Al$^+$ metal cations by radiative association}
\authorrunning{Bulut et al.}
\author{Niyazi Bulut \inst{1,2}
  \and
  Octavio Roncero\inst{3}$^*$
  \and
  Mar{\'i}a Mallo\inst{3}
  \and
  Germ\'an Molpeceres\inst{3}
  \and
  Marcelino Ag\'undez\inst{3}
  \and
  Maysa Yusef-Buey\inst{3}
  \and
  Jos\'e Cernicharo\inst{3}
  \and
  Carlos Cabezas\inst{3}
  \and
  Piotr S. Zuchowski\inst{4}
}
\institute{
       University of Firat, Department of Physics, 23169
           \and
  Faculty of Applied Physics and Mathematics, Gda\'nsk University of Technology, Gda\'nsk (Poland)
         \and
  Instituto de F{\'\i}sica Fundamental (IFF-CSIC), Serrano 123, 28006 Madrid (Spain)  \email{octavio.roncero@csic.es}
          \and
      Institute of Physics, Faculty of Physics, Astronomy and Informatics, Nicolaus Copernicus University in Torun, 
Grudziadzka 5, 87-100 Torun, Poland
}
\titlerunning{MC$_6$H$^+$ radiative association rate coefficients}
\abstract
{The Mg-bearing cations MgC$_4$H$^+$, MgC$_6$H$^+$, MgC$_3$N$^+$, and MgC$_5$N$^+$ have been recently observed in the carbon-rich envelope IRC\,+10216. These species are thought to form upon radiative association between Mg$^+$ and the corresponding neutral radical.
}
{This work aims to determine the radiative association rate coefficients for the cations Al$^+$, Mg$^+$, and Na$^+$ reacting with carbon chain radicals to estimate the relative importance of metal-bearing carbon cations for each of these metals.
}
{For this purpose we use statistical methods in combination with highly accurate ab initio calculations to  obtain the radiative association rate coefficient. Moreover, anisotropic effects beyond mono-dimensional capture model are considered, providing a more reliable description of the simulated processes.}
{The radiative association rate coefficients to form Mg-, Al- and Na-C$_6$H$^+$, as well as MgC$_4$H$^+$, MgC$_3$N$^+$, and MgC$_5$N$^+$, are calculated with the new restrictions considered in this work. These new rate coefficients are included in a chemical model of the C-rich AGB envelope IRC\,+10216.
}
{The abundances of MgC$_4$H$^+$, MgC$_6$H$^+$, and MgC$_5$N$^+$ are consistent with observations, with differences no greater than one order of magnitude, corroborating radiative association as a plausible formation mechanism for these adducts. However, the calculated abundance of MgC$_3$N$^+$ is about two orders of magnitude lower than observed, which indicates that either the chemistry of this species is significantly different from the others or the calculated rate coefficient is too low. The abundances obtained for 
Na-C$_6$H$^+$  and Al-C$_6$H$^+$ are around two orders of magnitude lower than for Mg, making very difficult to detect the analogous metal-bearing cations in IRC\,+10216.
}

  \maketitle

\section{Introduction}

Metal-bearing molecules are not observed in cold dense interstellar clouds \citep{Turner1991,Turner2005} because in these regions metals are expected to be largely trapped in dust grains. At the exception of some recent detections in very hot massive star-forming regions \citep{Ginsburg2019,Tachibana2019}, metal-bearing molecules are exclusively detected in envelopes around evolved stars. The high temperatures prevailing in the inner circumstellar regions around Asymptotic Giant Branch (AGB) allow metals to survive in the gas phase, and a fraction of them forms molecules, typically salts such as NaCl, KCl, AlCl, and AlF in carbon-rich envelopes \citep{Cernicharo-Guelin:87} and oxides such as TiO, TiO$_2$, AlO, and AlOH in oxygen-rich ones \citep{Tenenbaum2009,Tenenbaum2010,Kaminski2013}.

Most interestingly, a wide variety of organometallic molecules have been found in the cool outer layers of the C-rich AGB envelope IRC\,+10216 \citep{Guelin1993}. The existence of such molecules is not related to a high temperature, but to the gas-phase survival of metal atoms in the outer layers \citep{Mauron-Huggins:10} and their involvement in an exotic chemistry, first elucidated by \cite{Petrie1996}. The list of organometallic molecules detected in IRC\,+10216 and thought to be formed in the outer envelope is large and comprises MgNC \citep{Guelin1986,Kawaguchi1993}, MgCN \citep{Ziurys1995}, HMgNC \citep{Cabezas2013}, MgC$_2$ \citep{Changala2022}, MgC$_2$H \citep{Agundez2014,Cernicharo2019}, MgC$_4$H and MgC$_3$N  \citep{Cernicharo2019}, MgC$_6$H and MgC$_5$N \citep{Pardo2021}, HMgC$_3$N, NaC$_3$N \citep{Cabezas2023}, AlNC \citep{Ziurys2002}, CaC$_2$ \citep{Gupta2024}, CaNC \citep{Cernicharo2019_canc}, FeC \citep{Koelemay2023}, and FeCN \citep{Zack2011}.

The formation mechanism of these molecules is thought to rely on the availability of atomic metal cations. Such cations were initially considered chemically inert, but experiments by \cite{Smith-etal:83} indicated that Na$^+$ cations could rapidly form molecules via radiative association (RA), an idea originally proposed by \cite{Kirby-Dalgarno:78}. Later on, \cite{Petrie1996} proposed that MgNC could be formed in a two step process consisting on the radiative association of Mg$^+$ ions with large cyanopolyynes, followed by the dissociative recombination of the cationic complex with electrons, and the subsequent formation of neutral fragments. RA rate coefficients were systematically calculated for cations Na$^+$, Mg$^+$, Al$^+$, and Ca$^+$ reacting with cyanopolynes and polyynes \citep{Dunbar-Petrie:02,Petrie2004}, and these rate coefficients have been extensively used in chemical models aiming at explaining the formation of metal-bearing molecules in the outer envelope of IRC\,+10216 \citep{Millar2008,Millar2024,Cordiner2009,Cabezas2013,Cabezas2023,Cernicharo2019,Cernicharo2019_canc,Pardo2021}.

Recently, strong support to the mechanism proposed by \cite{Petrie1996} was provided by the detection in IRC\,+10216 of the cations MgC$_4$H$^+$, MgC$_3$N$^+$, MgC$_6$H$^+$, and MgC$_5$N$^+$ \citep{Cernicharo2023}. Unlike in the case of the neutral organometallics mentioned above, these cationic species are thought to arise directly in the radiative association of Mg$^+$ and the corresponding neutral radical.
{ Their detection therefore offers a way to check the link between
   RA calculated rate coefficients against observed abundances,
    without the need to involve the subsequent step of dissociative recombination,
    where branching ratios of the different fragmentation channels remain poorly known.
}

In \cite{Cernicharo2023} a statistical approach was used to calculate RA rate coefficients for Mg$^+$($^2$S) reacting with C$_4$H, C$_6$H, C$_3$N, and C$_5$N. The aim of this work is to extend the study to the atomic metal cations Na$^+$($^1$S), Mg$^+$($^2$S), and Al$^+$($^1$S), paying special attention to the role of the electronic structure of the system. We focus on the C$_6$H radical, whose ground electronic state is $^2\Pi$, while the corresponding adduct, MgC$_6$H$^+$, is of $^1\Sigma$ character. Since the ground electronic state of Mg$^+$ is $^2$S state, this implies an electronic transition. Polyyne radicals C$_n$H, widely detected in space, are linear species with two low-lying electronic states $^2\Sigma$ and $^2\Pi$, whose splitting varies with the length of the chain \citep{Taylor-etal:98,Garand-etal:10}. For $n$= 2, 3 and 4, the ground state is $^2\Sigma$, and the splitting with the excited $^2\Pi$ reduces gradually to only 213 cm$^{-1}$ for C$_4$H. For C$_6$H, the ground state is $^2\Pi$, which is 1413 cm$^{-1}$ (0.175 eV) below the
$^2\Sigma$ state \citep{Taylor-etal:98,Garand-etal:10}.

\section{Electronic states crossing}

\begin{figure}
    \resizebox{\hsize}{!}{\includegraphics[width=\linewidth]{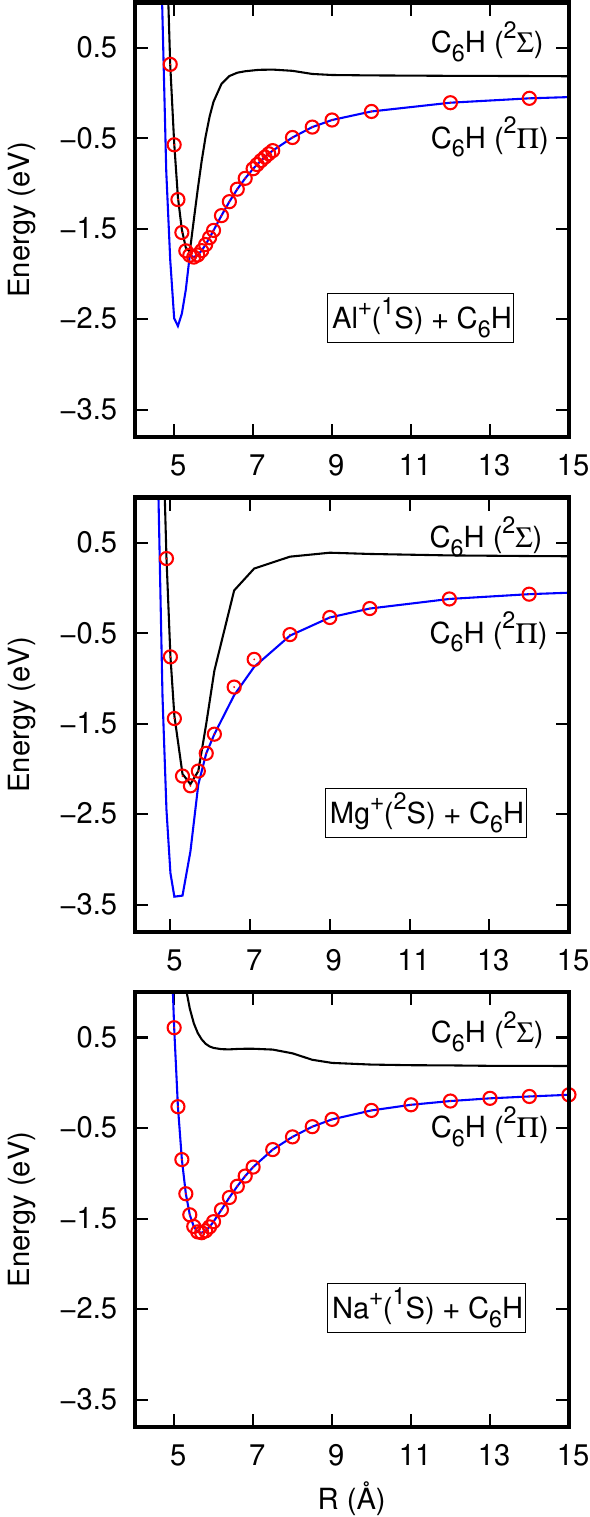}}
    \caption{Potential energies of the three lower electronic states of M$^+$ + C$_6$H (M$^+$= Na$^+$, Mg$^+$ and Al$^+$)
      for a collinear geometry. Electronic energies have been obtained with a MRCI method, using the MOLPRO\citep{MOLPRO-WIREs} package,
      as a function of the distance $R$ of M$^+$ to the center-of-mass of C$_6$H in \AA.
      {Lines represent the first two A' electronic energies corresponding to a $\Sigma$ and $\Pi$ state,
    which cross for Mg$^+$ and Al$^+$. Open circles show the first A'' electronic energy, corresponding to a $\Pi$
    state, which is { numerically  degenerate (with some error in the case of Mg)} with that obtained in the A' representation, thus giving direct evidence of
    the crossing at this linear configuration.}
    }
    \label{fig:mrci-collinear-pes}
\end{figure}

The three lower electronic states of the system C$_6$H+M$^+$ (M=Na, Mg and Al) were calculated
using the internally contracted multi-reference configuration interaction (ic-MRCI)
method  \citep{Werner-Knowles:88,Werner-Knowles:88b}, implemented in the MOLPRO suite of
programs \citep{MOLPRO-WIREs}. Calculations were carried out for a collinear geometry
as a function of the distance between M$^+$ and the center of mass of C$_6$H,
starting at 15 \AA~ down to the repulsive region of the potential, at approximately 2 \AA.
In these calculations the molecular orbitals are optimized using  a
state-averaged complete active space self-consistent field (SA-CASSCF) method,
with 6 electrons in an active space of 5 orbitals in $C_s$ symmetry using the aug-cc-pVTZ (aVTZ)
basis of \cite{Dunning:89}. The active space includes two orbitals of $\sigma$
symmetry (valence orbital in the case of Mg and terminal C atom) and three lowest $\pi$ orbitals
of C$_6$H. Three electronic states were converged for the system, in Fig.~\ref{fig:mrci-collinear-pes},
two A' (lines) and one A'' (points), associated to $\Sigma$ and $\Pi$ states in the linear equilibrium geometry.
{ The long-range interactions have been evaluated within perturbation theory \citep{Hapka-etal:12} using the MATROP utility
  in MOLPRO package, by using the isolated  wave functions of M$^+$ and C$_6H$ and calculating the electrostatic
  interaction as a function of their distance, in the interval $20 \leq R\leq 50$\AA.
  The energies were fitted to $A/R^2$ (see Eq.~(1)), with $A$=59.31 eV \AA$^2$ ,
  corresponding to an electric dipole moment for C$_6$H($^2\Pi$)  of 5.54 Debye.
   It is also found that the electric
      dipole of the excited $^2\Sigma$ state is
      much smaller than that of the $^2\Pi$ state,
      what explains why the interaction potential of the excited
      electronic state is visibly  weaker and flat in the asymptotic region,
      as can be seen in Fig. \ref{fig:mrci-collinear-pes}.
    }

In the ground $^2\Pi$ state of C$_6$H the interaction with the metal cation  at collinear geometries is attractive as expected in charge-electric dipole  interactions, presenting a relatively deep well of 1.6, 2.4 and 1.8 eV for Na$^+$, Mg$^+$ and Al$^+$, respectively. The situation is rather different in the excited $^2\Sigma$ state of C$_6$H, whose interaction with the metal cations is rather constant, until reaching $\approx$ 7-9 \AA. At this distance, the interaction with Na$^+$ becomes repulsive, while with Mg$^+$ and Al$^+$ becomes suddenly very attractive, although Mg$^+$ presents a clear monotonic decrease of the electronic energy, while Al$^+$ shows a small activation barrier. Nevertheless, Mg$^+$ and Al$^+$ present crossings with the $^2\Pi$ state, thus giving rise to $^1\Sigma$ adducts. This explains the recent microwave observation of MgC$_6$H$^+$($^1\Sigma$) in IRC\,+10216 \citep{Cernicharo2023}.
 
\begin{figure}
\resizebox{\hsize}{!}{\includegraphics[width=\linewidth]{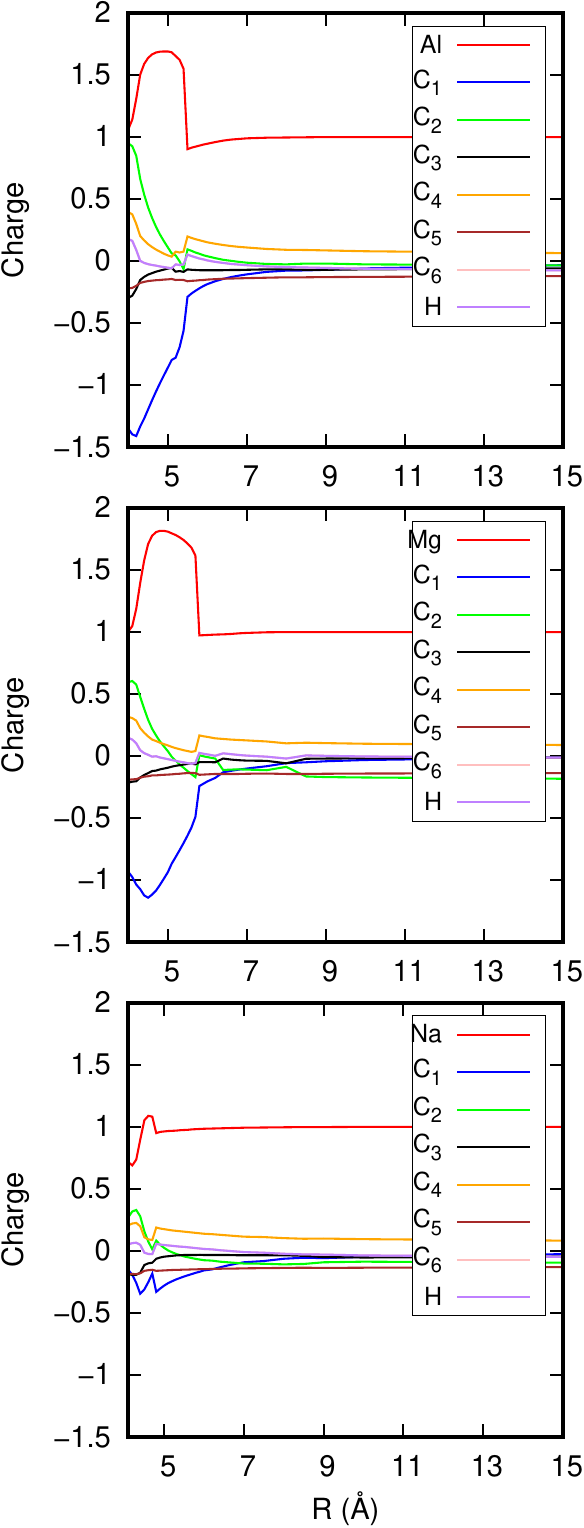}}
\caption{Atomic charges ( in units of the proton charge,$\approx$  1.602 $\times 10^{-19}$ { C})
  of M$^+$C$_6$H (M$^+$= Na$^+$, Mg$^+$ and Al$^+$)
  in the ground electronic state using the Mulliken method  as a function of
  the distance $R$ of M$^+$ to the center-of-mass of C$_6$H in \AA.
  $C_i$ denotes the six carbon atoms in a line, with $i$=1 being the closet to M$^+$ atom.}
\label{fig:mrci-collinear-charge}
\end{figure}

The sudden change of slope of the $^2\Sigma$ state occurring for Mg$^+$ and Al$^+$ can be attributed to an electronic crossing, with strong ionic character as M$^{2+}$ + C$_6$H$^-$. To check this, the atomic charges on the ground electronic state were calculated for the three systems, using Mulliken decomposition \citep{Mulliken:55}, and are shown in Fig.~\ref{fig:mrci-collinear-charge}. The charge on the metallic ions M$^+$ remains singly positive until $R$=5-7 \AA, where Al$^+$ and Mg$^+$ show a sudden change up to a value larger than 1.6 while Na$^+$ charge remains approximately +1. These sudden changes occur at the crossing, showing the  M$^{2+}$ ionic character of this state, whose energy decreases much rapidly than the others in the interval R=5-7 \AA.  The excess of negative charge is mostly in C$_1$, the closest carbon atom  to the metal cation.

The situation can be further rationalized in terms of ionization energy progression of the ions \citep{NIST_AtomicSpectraDatabase}, which are 47.29, 15.04 and 18.83 eV for Na$^+$, Mg$^+$ and Al$^+$, respectively. Clearly, the closed shell Na$^+$ has a much higher ionization energy, shifting to considerably higher energy the M$^{2+}$ + C$_6$H$^-$. Mg$^+$ and Al$^+$ cations have one or two 3s electrons, much easier { transferred} to form the M$^{2+}$ -C$_6$H$^-$ ionic structure
  
\section{M$^+$C$_6$H isomer exploration}

\begin{figure}
\resizebox{\hsize}{!}{
  \includegraphics[width=\linewidth]{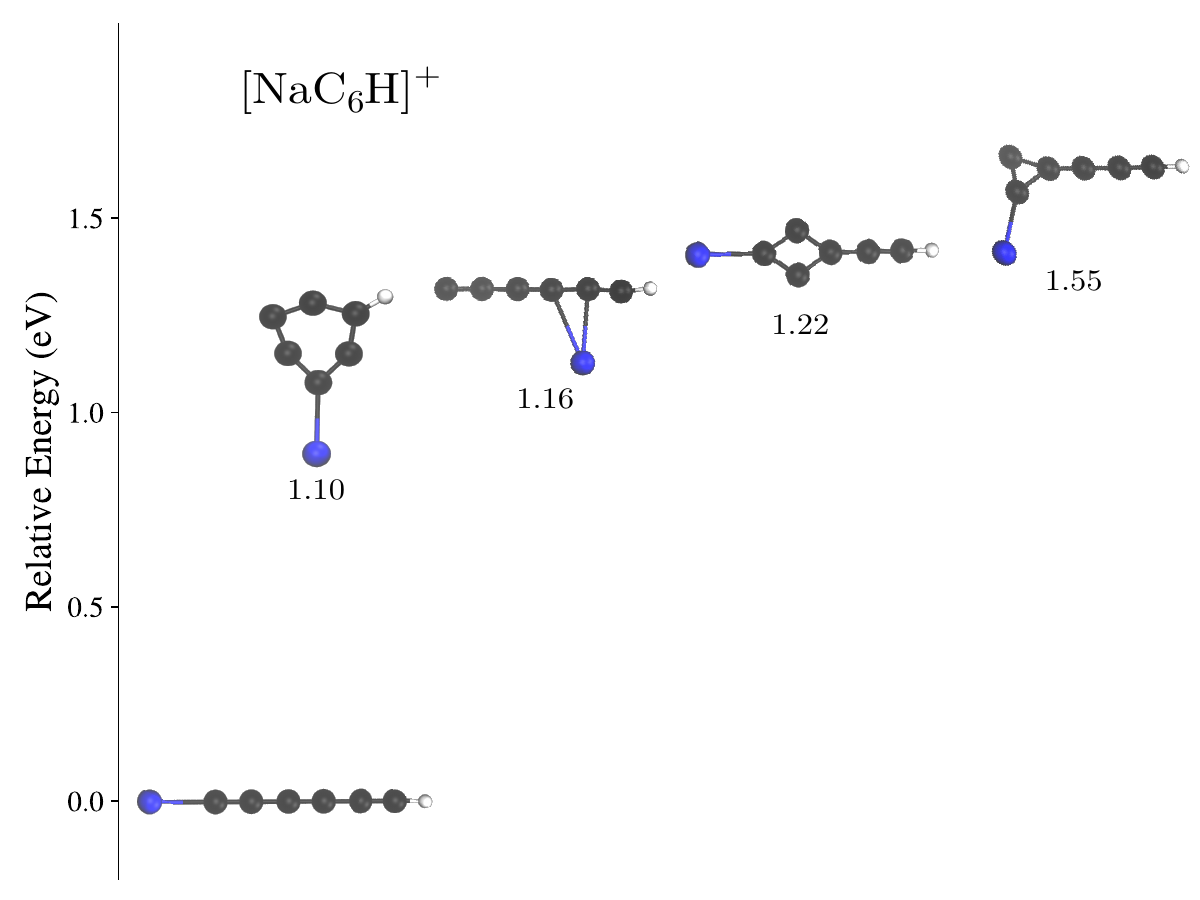}}

\resizebox{\hsize}{!}{
  \includegraphics[width=\linewidth]{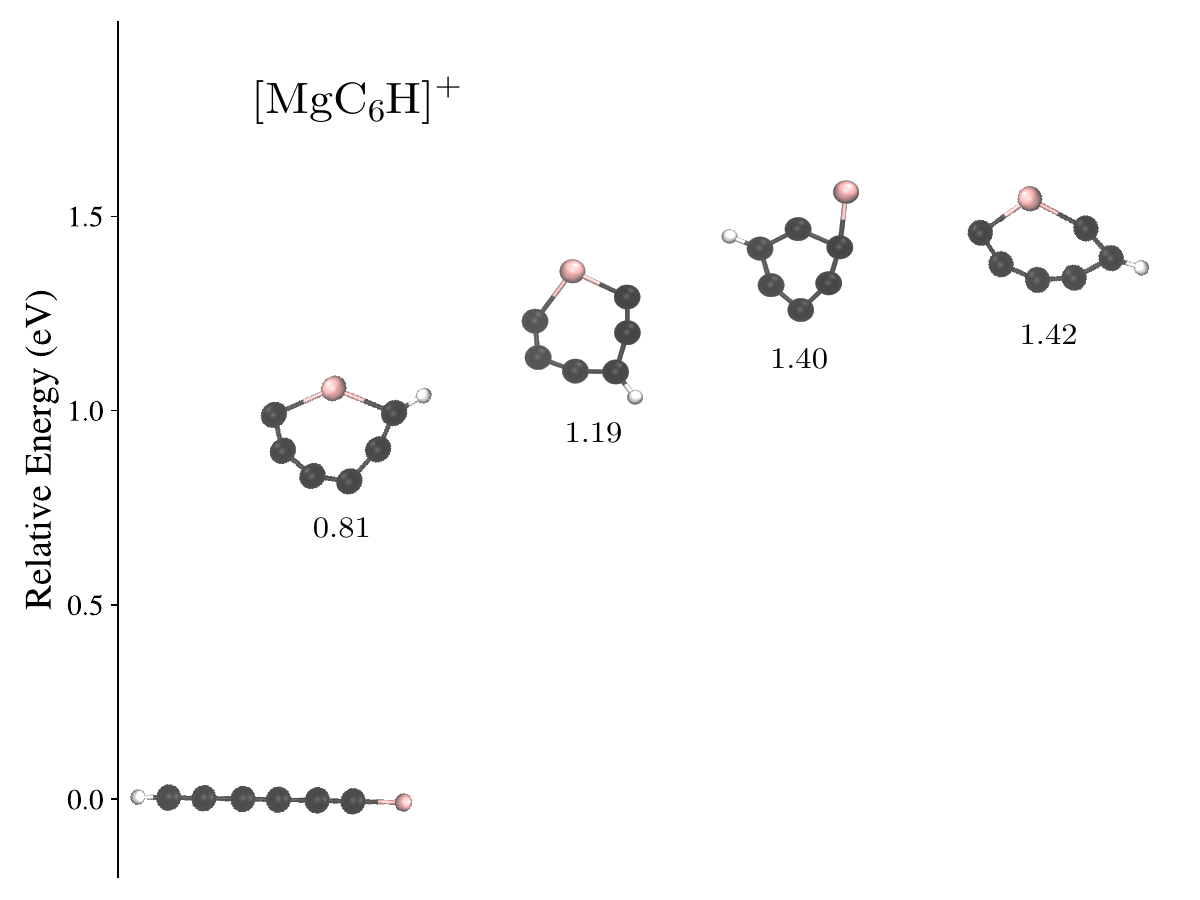}}
\resizebox{\hsize}{!}{
\includegraphics[width=\linewidth]{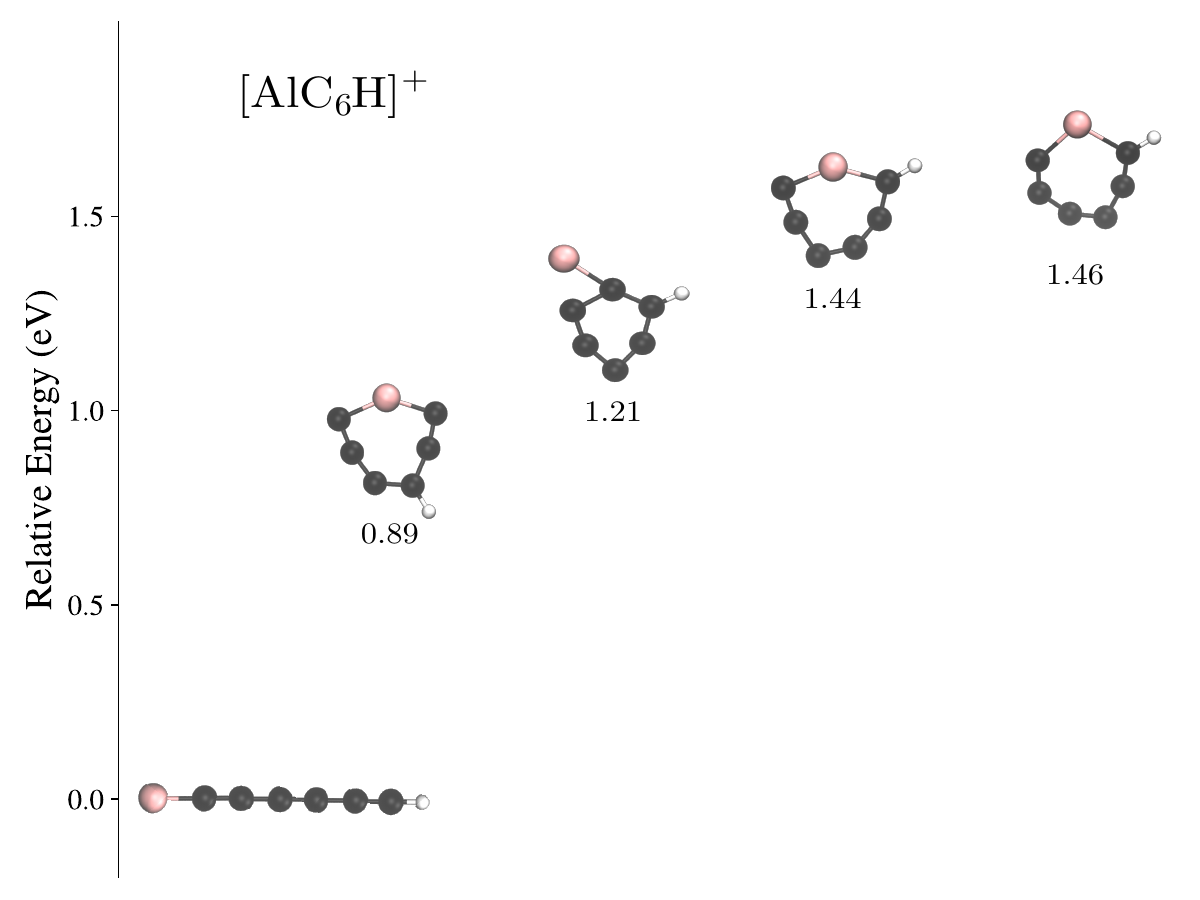}}
\caption{Lowest five isomers of M$^+$C$_6$H computed with automatic reaction search exploration.}
\label{fig:isomers-scheme}
\end{figure}

After association of the metal cation with the C$_6$H substructure, a natural question arises: can the resulting complex undergo isomerization away from the linear geometry, thereby slowing down or even preventing radiative association (RA) in the linear adduct? We confirmed that the formation of the linear metal-adducts is the most favorable reaction pathway by checking the isomeric landscape for \ce{MC6H+}. The investigation of the minima in the complex PES of the adducts is too complicated and requires of several thousands of single point calculations to be evaluated. For that reason, in this part of the work we opted to use density functional theory (DFT) calculations in combination with automatic reaction discovery techniques to investigate the isomeric panorama. 

For the automatic exploration of the isomeric landscape we employ the automatic force induced reaction (AFIR) method \citep{maeda_implementation_2018, maeda_systematic_2013, maeda_toward_2023} as implemented in the \textsc{Grrm23} code \citep{grrm23}. The protocol closely follows that used in \citet{basilio_double_2026}, although with some modifications to the energetic and search parameters. Briefly, a single AFIR function is applied starting from the linear configurations of each metal complex, \ce{Al+}, \ce{Mg+}, and \ce{Na+}, to search for energetically competitive isomers within the configurational space. The AFIR function is applied with a collision factor of $\gamma$ = 4.14 eV. {This energy can be interpreted as an estimate of the highest energy barrier that the AFIR method can overcome. Therefore, it is deliberately chosen to be sufficiently large to allow the algorithm to escape from the highly stable linear minimum that serves as the initial structure in the simulation.} The Monte Carlo sampling of the intramolecular fragments on which the AFIR function acts is selected randomly from the pool of previously discovered structures using a Maxwell-Boltzmann distribution at a temperature of 3000 K, corresponding to the default settings of \textsc{Grrm23}. {This parameter does not represent a physical temperature. Instead, it determines the probability with which a given isomer is selected as the starting structure in an AFIR search. As will be shown later, the minimum energy difference between the linear isomers and all other structures exceeds 0.8 eV (approximately 9000 K). Consequently, most AFIR calculations will be initiated from the linear isomer, which is consistent with our objective of identifying low-energy isomers that can compete with this structure.} As in our recent investigation \citep{basilio_double_2026}, the search is restricted to minima on the electronic potential energy surface, since including transition states connecting them would significantly increase the computational cost. In addition, a stopping criterion is imposed whereby the search is terminated when the lowest 16 identified isomers remain unchanged after the last 32 search attempts. The electronic structure calculations performed during the exploration of the isomeric PES are carried out with the \textsc{Gaussian16} software using the TPSSh exchange and correlation functional \citep{tao_climbing_2003, perdew_workhorse_2009} in combination with the def2-SVP basis set \citep{Weigend2005}. We improve the results of our search with two refinements. In the first place we improve the molecular geometries with geometry optimizations with a larger basis set, def2-TZVPD \citep{rappoport_property-optimized_2010}. The geometries and molecular Hessians are therefore calculated at the TPSSh/def2-TZVPD level \citep{rappoport_property-optimized_2010}. Finally, electronic energies are corrected at the DLPNO-CCSD(T)/cc-pVTZ level \citep[][in addition to the previous citations]{guo_communication_2018, Dunning:89}.  From our search we obtained an average per molecular formula of 34 isomers, for which we only show in Figure \ref{fig:isomers-scheme} the lowest five ones. Our calculations unambiguously show that the linear geometry is the most
stable one for the three metals under consideration, with the rest of the isomers at a minimum energy separation of 0.8 eV (in the case of \ce{Mg+}) and a maximum of 0.9 eV (in the case of \ce{Al+}). In all cases, the second and subsequent isomers lie within a relatively narrow energy range, with the energy difference between the second and fifth isomers being approximately 0.5 eV, which is smaller than the separation between the first and second isomers. Therefore, from our exploration of the isomeric landscape of \ce{MC6H+}, we can conclude that the linear isomer is the most stable and, most likely, the one formed preferentially in the \ce{M+ + C6H} reaction. Together with the finding that bimolecular H-elimination is not competitive, this supports radiative association as a major pathway for the formation of the title molecules. 

\section{Radiative association rate coefficients}

After combining the results from Sections 2 and 3, and considering that H-elimination is endothermic, the only possible channels are either RA from the association adduct or the M$^+$ + C$_6$H back-dissociation. The adduct is directly accessible from the M$^+$ + C$_6$H channel, as seen in Fig.~\ref{fig:mrci-collinear-pes}, by the strong attractive ion-electric dipole interaction
\begin{eqnarray}\label{eq:ion-dipole-interaction}
  V_1= -{q_{ion} {\bf d}\over 4\pi\epsilon_0} {\cos\theta\over R^2},
\end{eqnarray}
where ${\bf d}$ is the electric dipole of the molecule, $q_{ion}$ is
the charge of the ion, ${\bf R}$ is the vector joining the center-of-mass of A to the M$^+$ ion (of norm $R$),
with $\cos\theta={\bf R}\cdot{\bf d}/ R d$. The linear C$_6$H molecule has a rather large electric dipole along the molecular axis, favoring the linear approach between the two reactants.
The same {  long range charge-electric dipole  interaction are valid} for the three metal cations, all governed by the C$_6$H electric dipole.

Once the (M-C$_6$H$^+$$)$$^*$ complex is formed, it can either dissociate back, with a rate coefficient $K_B$, or emit
a photon to stabilize the MC$_6$H$^+$ ion, with a radiative constant $K_E$. Thus the total radiative association
rate coefficient, $K_{RA}$, is obtained as \citep{Herbst:76,Bass-etal:81,Herbst:87,Klippenstein-etal:96}
\begin{eqnarray}
  K_{RA} (T) = K_F(T)  {K_E(T)\over{K_E(T)+K_B(T)}},
\end{eqnarray}
where the formation rate, $K_F(T)$, is rather fast, specially at low temperature, because of the long-range interactions, and the rate coefficient $K_{RA}$ depends strongly on the ratio $R(T)={K_E(T)/({K_E(T)+K_B(T)})}$, which is very sensitive
to the molecular properties.

$K_F(T),K_B(T)$ and $K_E(T)$ are calculated here using statistical methods with a home made program following
the methods described previously \citep{Herbst:76,Bass-etal:81,Miller:87,Klippenstein-etal:96,Dunbar-Petrie:02,Cernicharo2023}. The microcanonical rate coefficient for each individual process is first obtained as a function of the collisional energy, $E$. The thermal rate coefficients are obtained by averaging over energy using a Boltzmann distribution.

For the statistical calculations presented below, the minima of the three MC$_6$H$^+$ and C$_6$H 
are determined and their normal modes calculated with the CCSD(T) method with the  aug-cc-pVTZ
(aVTZ) basis of \cite{Dunning:89}, using the MOLPRO-12 suite of programs \citep{MOLPRO-WIREs}. 

\subsection{Capture}

The strong ion-dipole interaction in Eq.\,(\ref{eq:ion-dipole-interaction}) rapidly { reorients} the system to a collinear geometry, and hence a simple capture model based on the monodimensional radial coordinate can be applied \citep{Chesnavich-etal:80}, similar
to the Langevin model \citep{Levine-Bernstein:87}. The effective potential, the sum of the charge-electric dipole interaction and the rotational energy, is given by
\begin{eqnarray}
    V_{eff}(R,J)=\left\lbrack {\hbar^2\over 2\mu} J(J+1) - C\right\rbrack
\end{eqnarray}
with $J=\mu v b$ being the total angular momentum, $\mu=m_{M^+} m_{C_6H}/(m_{M^+}+ m_{C_6H})$  the reduced mass and $C={q_{ion} {\bf d}/ 4\pi\epsilon_0} $. The effective potential is attractive for $J < \sqrt{2 \mu A} /\hbar$, and repulsive elsewhere, giving a maximum classical impact parameter of
\begin{eqnarray}\label{eq:capture-xsection}
b^2={C\over \hbar^2 E} \quad\longrightarrow \sigma(E) = {\pi C\over \hbar^2 E}.
\end{eqnarray}
The integration of the capture cross section with a Boltzmann distribution can be done analytically \citep{Gradshteyn-Ryzhik:80}, obtaining the thermal rate coefficient of the form \citep{Moran-Hamill:63}
\begin{eqnarray}\label{eq:1d-capture}
  K_F(T)= q_e \sqrt{ {8 \pi C^2 \over \mu k_B T}},
\end{eqnarray}
where $q_e=$ 1/2, 1/8, and 1 for Al$^+$, Mg$^+$, and Na$^+$, respectively, is the electronic partition function. 
\begin{figure}
\resizebox{\hsize}{!}{\includegraphics[width=0.75\linewidth]{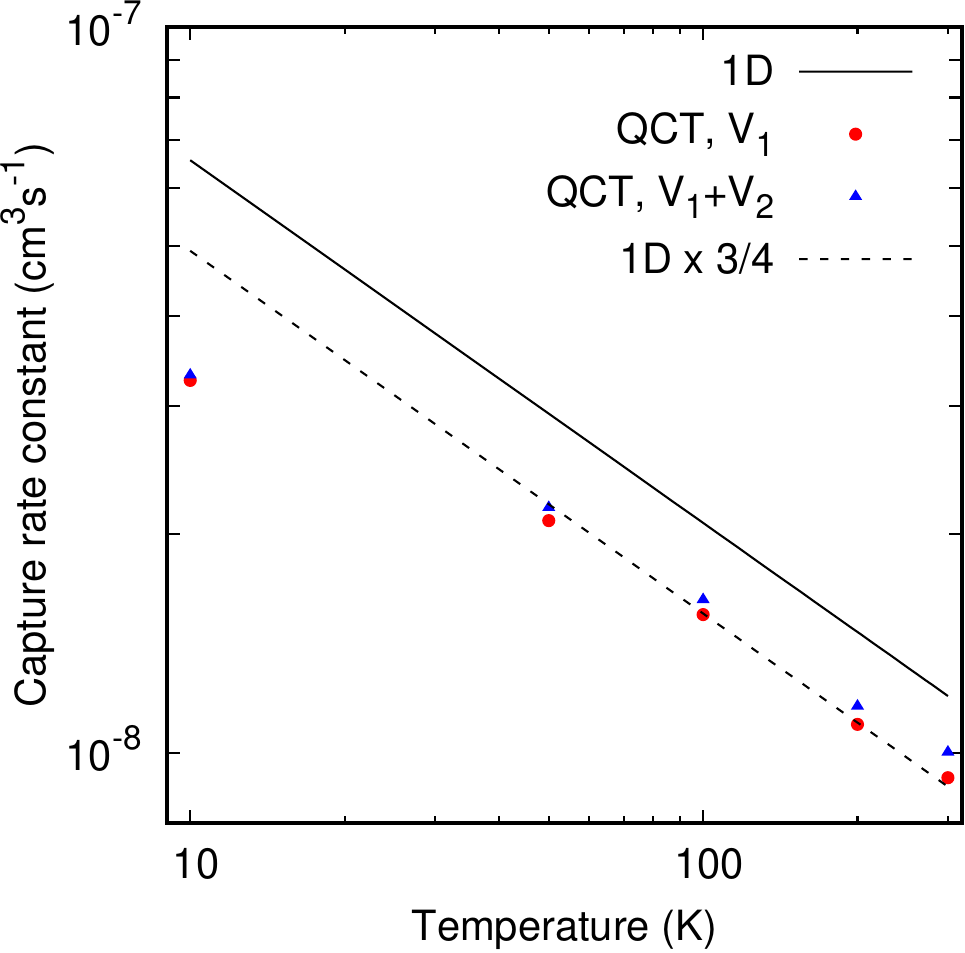}}
\caption{Capture rates, considering the $K_F$, Eq.\,(\ref{eq:1d-capture}),
  analytical one-dimension model (1D black line), and the quasi-classical results (QCT)
  using two long-range model potentials, V$_1$ in Eq.\,(\ref{eq:ion-dipole-interaction}),
  and V$_2$, in Eq.\,(\ref{eq:charge-induced-dipole}). The 1D $\times$ 3/4 corresponds $K_F \times c$ ,
  where $c$=3/4 is the locking parameter $c$=3/4, constant in temperature in this case.}
\label{fig:captureK}
\end{figure}

This approach represents the locked dipole model of \cite{Moran-Hamill:63} and it often overestimates the rate coefficient at low temperature.
\cite{Su-Bowers:73,Su-Bowers:75} proposed the average dipole orientation model, introducing the locking parameter (here denoted as $c$)
to reproduce how the dipole aligns with the ion field.
Another alternative is the use of the adiabatic capture method (ACM) \citep{Clary:85,Quack-Troe:74}, in which the dependence of the 
orientation between the linear C$_6$H and the ion is considered, including rotational levels of the linear C$_6$H. This treatment becomes very demanding in this case because of the very small C$_6$H rotational constant. One alternative is to neglect the Coriolis couplings in the centrifugal sudden approach (CS), as done by \cite{Clary:85}. However, the CS-ACM method overestimates the rotational constants, and therefore it is expected to underestimate the capture rate coefficient.

Instead, here we use a quasi-classical method (QCT), described in Appendix~\ref{appendix:qct}, as done previously by \cite{Su-Chesnavich:82,Chesnavich-etal:80}. The QCT capture rate coefficients, obtained for two model long-range  potentials, are shown in Fig.~\ref{fig:captureK} and compared to the one-dimension $K_F$ rate of Eq.\,(\ref{eq:1d-capture}). The QCT results are lower than the  $K_F$ rate, and the two model long-range potentials yield approximately the same results. The locking parameter $c$ slightly depends on temperature, growing larger as temperature decrease. For the temperature interval of interest, between 100 and 300 K, $c$=3/4. 

\subsection{Back dissociation and emission rates}

The back dissociation microcanonical rate coefficient takes the simple form \citep{Miller:87}
\begin{eqnarray}
  K_B(E,J)={\sum_{J_A,\ell} P(E,J_A,\ell)\over h \rho_{MA^+}(E,J)}
\end{eqnarray}
where $J_A$ and $\ell$ are the angular momenta associated to molecule A and of M$^+$
with respect to A=C$_6$H (with ${\bf J}={\bf J}_A+\boldmath{\ell}$). The cumulative probabilities are obtained as in the capture process as
\begin{eqnarray}
  P(E,J_A,\ell)=\left\lbrace
  \begin{array}{c}
    0 \quad E < E_{J_A}\\
    0 \quad \ell > \sqrt{2\mu C}/\hbar\\
    1 \quad {\rm elsewhere}.
  \end{array}
  \right.
\end{eqnarray}
Finally, the density of states of MA$^+$, $\rho_{MA^+}$, depends on the binding energy
and the frequency of the normal modes, and is calculated using the method of \cite{Stein-Rabinovitch:73}.

The radiative rate coefficient is given by \citep{Herbst:82}
\begin{eqnarray}\label{emision-rate}
  K_E(E,J) =\sum_N P_N(E,J) A^J_N,
\end{eqnarray}
where $N={n_1, n_2, ... j, n_M}$ is a collective vibrational quantum number
specifying the vibrational states of the $n_M$ normal modes of  (MA$^+$$)$$^*$,
$ P_N(E,J)$ is the probability distribution of level $N$  in an energy interval around $E$. $A^J_N$ is the
radiative rate of level $N$ corresponding to the sum of the Einstein coefficients
over all possible single quantum emission of all the normal modes.
Using the rigid rotor approximation and performing the summation over the H\"onl-London
factors \citep{Whiting-Nicholls:74,Whiting-etal:80} the vibrational emission rate does no longer
depend on $J$ and can  be
defined as
\begin{eqnarray}
  A_N\approx\sum_{N'}= {1\over 3\pi\epsilon_0 \hbar^4}\left( {E_N-E_{N'}\over c}\right)^3
  \vert M_{NN'}\vert^2,
\end{eqnarray}
where the  matrix elements in the normal mode approximation can be written
as
\begin{eqnarray}
  M_{NN'} 
  = \sum_{i=1}^{n_M} \sum_{\beta=x,y,z} {\partial d_\beta\over \partial Q_i}
  \sqrt{n_I\hbar \over \mu\omega_i}\quad \delta_{n_i,n_{i'-1}}\prod_{j\neq i} \delta_{n_j,n_j'}
\end{eqnarray}
where only one vibrational quantum transition is considered because the
electric dipole component, $d_\alpha$, is expanded around equilibrium, $Q_i$=0,
as
\begin{eqnarray}\label{dipole-expansion}
  d_\beta = d^0_\beta + \sum_i Q_i \left.{\partial d_\beta\over \partial Q_i}\right\vert_{Q_i=0} .
\end{eqnarray}
The dipole moments are calculated along each normal mode, $Q_i$, of the MC$_6$H$^+$ adduct with the finite field method at CCSD(T) level, by including homogeneous electric fields for  each cartesian coordinate, $\beta$. 

In Eq.\,(\ref{emision-rate}) the number of vibrational states is very high, and the complete sum over all quantum energy at energy $E$ is substituted by a Monte Carlo sampling. The number of samples is varied, obtaining a good convergence at values of 10000\,$\times$\,$n_M$ per energy.
\begin{figure}
    \resizebox{\hsize}{!}{\includegraphics[width=\linewidth]{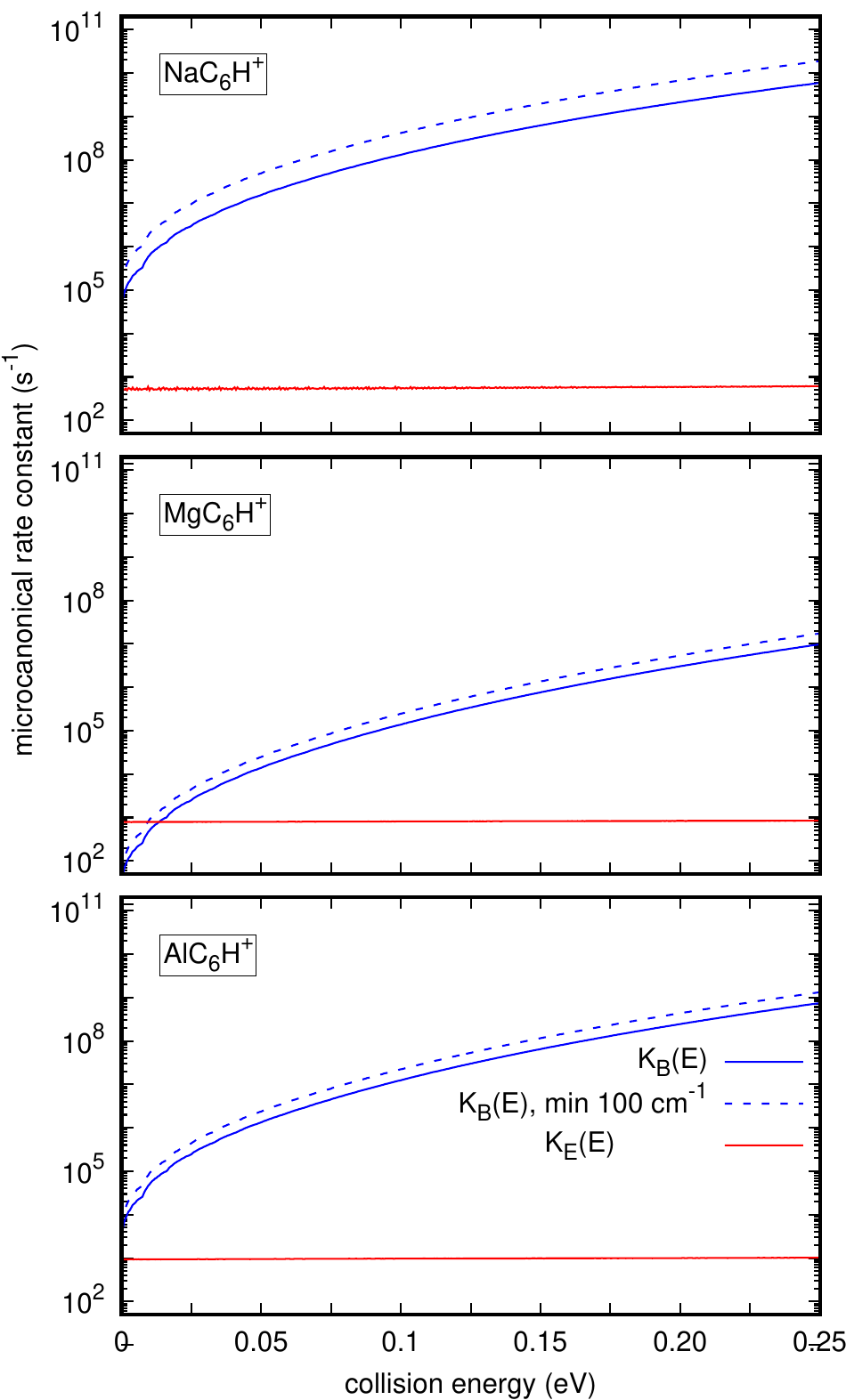}}
    \caption{Backward dissociation and emission microcanonical rate coefficients for the three MC$_6$H$^+$ systems, as indicate in each panel. Two cases are shown for the backward dissociation (in blue): solid lines using the normal mode frequencies of Table~\ref{tab:NM}, dashed lines setting 100 cm$^{-1}$ as the minimum of the normal mode frequencies.}
    \label{fig:microK}
\end{figure}

The micro-canonical rate coefficients are shown in Fig.~\ref{fig:microK}. The emission rates are nearly constant as a function of energy and very similar for the three systems, of the order of 10$^3$ s$^{-1}$. On the contrary, the backward dissociation rates show a monotonically increasing behavior with temperature, with very similar slope in the three systems because the normal mode frequencies are rather similar. However, the limit at zero collision energy differ up to three orders of magnitude, with MgC$_6$H$^+$ being the slowest, 10$^2$ s$^{-1}$,
and NaC$_6$H$^+$ being the fastest,  10$^5$ s$^{-1}$. These values are inversely proportional to the binding energies, being 3.5, 2.5 and 1.5 eV for Mg, Al, and Na, respectively, as shown in Fig.~\ref{fig:mrci-collinear-pes}. This clearly indicates that at low energy $K_E$ is faster than back dissociation for Mg, while this situation reverts for Al and Na.

The lower normal mode frequencies are those that contribute more to the density of states, $\rho_{MA^+}(E,J)$, slowing down the dissociation rate. These low frequencies have larger errors since they are more coupled to rotations, and the absolute errors associated with a numerical derivation of the vibrational frequencies penalize low frequencies more. In condensed phase, these low frequencies are sometimes constraint to a minimum value to reduce their effect \citep{Ribeiro-etal:11,Velmiskina-etal:25}. For linear polyatomic systems in gas phase, low frequency normal modes can also be associated to quasi-rotational states, and can be considered to be rather uncoupled to the rest of vibrational states.

\begin{table}
\caption{Normal mode frequencies (in cm$^{-1}$) and electric dipole derivatives (in atomic units) obtained.}
\label{tab:NM}
    \centering
    \begin{tabular}{|cc|cc|cc|}
  \hline
       \multicolumn{2}{|c|}{  Na}   &   \multicolumn{2}{c|}{Mg }    &     \multicolumn{2}{c|}{Al} \\
\hline
   $E_N$    & $\partial d /\partial Q_N$  &  $E_N$    & $\partial d /\partial Q_N$       &  $E_N$    & $\partial d /\partial Q_N$ \\
\hline
      41.7  &   1e-4        &           56.4  & 0.002       &         55.5 & 0.002 \\
      41.7  &   1e-4         &          56.4  & 0.002        &        55.5 & 0.002 \\
      84.9  &  0.001     &         139.3  & 1e-5    &        137.3 & 5e-4 \\
      97.9  &  0.001      &        139.3  & 1e-5    &        137.3 &  5e-4 \\
      97.9  &  0.001     &         262.3  & 0.002      &        257.9 & 0.001 \\
     249.5  &  0.001      &         262.3  &0.002       &        257.9 &  0.001 \\
     249.5  &0.001      &         401.7  &0.001       &        420.3 &  6e-4 \\
     464.0  &4e-4     &         452.9  &  4e-4      &        461.5 & 4e-4 \\
     464.0  &4e-4     &         452.9  & 4e-4      &        461.5 & 4e-4 \\
     507.6  & 5e-4      &         486.2  &3e-4      &        493.0 & 4e-4 \\
     507.6  & 5e-4      &         486.2  &3e-4       &        493.0 &  4e-4 \\
     628.6  &0.001      &         671.3  &  0.005     &        690.4 & 0.004 \\
     681.0  & 0.004     &         671.3  &  0.003      &        690.4 & 0.004 \\
     681.0  & 0.004     &         749.8  & 0.002       &        797.1 &  0.002 \\
    1180.3  &0.001      &        1214.8  & 3e-4      &       1252.7 &  5e-4 \\
    2058.7  &0.001      &        2053.4  & 0.004       &       2040.0 &   0.007 \\
    2164.6  &0.008      &        2158.0  & 0.002      &       2152.1 & 0.005 \\
    2259.9  &0.008     &        2241.6  & 0.006      &       2216.5 &   0.010 \\
    3442.6  &0.009     &        3442.6  & 0.008      &       3431.2 &   0.009 \\
\hline    
\end{tabular}
\end{table}

In order to analyze the sensitivity of  K$_B$(E) to low-lying normal modes, here we consider a limit value for the low frequencies, 100 cm$^{-1}$ (dashed lines in Fig.~\ref{fig:microK}), producing changes of factors of approximately 2-4  at low temperatures, being larger for Na which shows a higher number of low frequencies. In light of this sensitivity analysis, we consider justified to rise the lower limit of the vibrational frequencies to 100 cm$^{-1}$.

It is worth noting that \cite{Cernicharo2023} used another electronic basis set, resulting in two degenerate imaginary normal modes only for the case of MgC$_6$H$^{+}$. These imaginary frequencies were transformed to a real value, larger than 250 cm$^ {-1}$. As a consequence, the backward dissociation rate obtained in that work was considerably  larger than in this work. As discussed below, such situation yielded lower radiative association rate coefficients. However, the inclusion of a lower frequency limit (of 100 cm$^{-1}$) and a locking factor (of 3/4) in the present treatment compensates that problem.

\begin{figure}
    \resizebox{\hsize}{!}{\includegraphics[width=\linewidth]{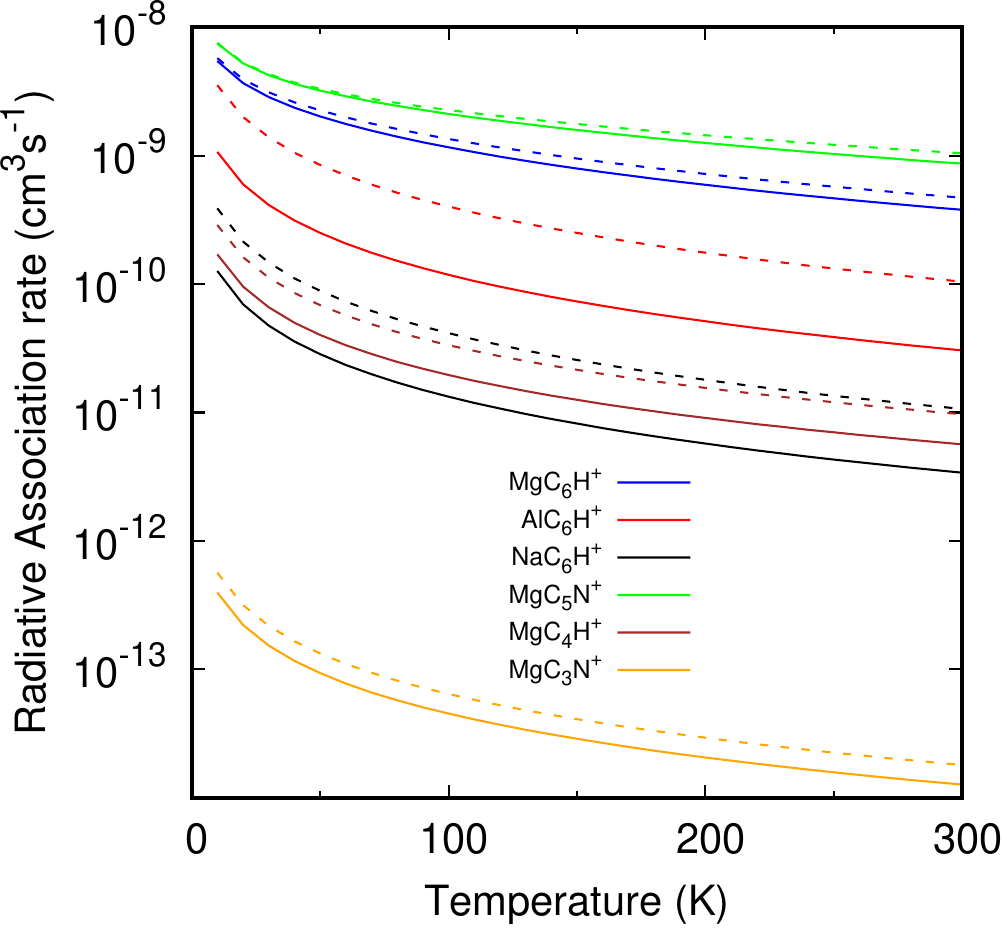}}
    \caption{Radiative association rates for MgC$_6$H$^+$, AlC$_6$H$^+$ and NaC$_6$H$^+$, MgC$_5$N$^+$, MgC$_4$H$^+$ and MgC$_3$N$^+$. Solid lines are obtained by limiting the normal modes frequencies to 100 cm$^{-1}$, dashed lines with the original frequencies.
    The locking factor c=3/4 has been included for all the systems.}
    \label{fig:RArates}
\end{figure}

The radiative association rates are calculated using these two limiting cases and   with same long range interaction, characterized by the isolated molecule electric dipole, including the common locking factor c=3/4, shown in Fig.~\ref{fig:RArates}. In addition, the radiative emission rates for the three systems are approximately the same. Therefore, all the differences are explained by the well depths and normal modes frequencies, translated into different back-dissociation rate coefficients.

The differences between the three metal cations are mostly explained by the progressive increase of the well depth in the sequence Na$^+$, Al$^+$ and Mg$^+$. 
Low normal modes frequencies have an enormous effect on the radiative association rate, by lowering the re-dissociation rate. This is the reason of the differences between the rates of MgC$_6$H$^+$ obtained here and those reported previously in \cite{Cernicharo2023}, in which the lower frequencies below 100 cm$^{-1}$ were not included. 

The recommended radiative association rate coefficients are listed in Table~\ref{tab:RArates}, corresponding to those calculated with a lower limit of 100 cm${^-1}$ for the normal mode frequencies. These rate coefficients are used below to analyze their impact in the molecular abundances of the different adducts considered in this work in the chemical model of IRC\,+10216 (see next section). In addition to the rates of MgC$_6$H$^+$, AlC$_6$H$^+$ and NaC$_6$H$^+$, we also include in Table ~\ref{tab:RArates} the rates for MgC$_5$N$^+$, MgC$_4$H$^+$ and MgC$_3$N$^+$, calculated with the same method. 

\begin{table}
    \caption{A and b parameters of the radiative association rate coefficients fitted to the expression $k_{RA}$(T) = A(cm$^3$ s$^{-1}$) ($T$/300 K)$^b$,
      for MgC$_6$H$^+$, AlC$_6$H$^+$, NaC$_6$H$^+$, MgC$_5$N$^+$, MgC$_4$H$^+$ and MgC$_3$N$^+$.
      These fits correspond to limiting the normal mode frequencies to 100 cm$^{-1}$,
      {\  and are appropriate in the 10-300 K temperature range.}
    }
\label{tab:RArates}
    \centering
    \begin{tabular}{|c|c|c|}
        \hline
       System & A(cm$^3$ s$^{-1}$) & b\\
        \hline      
       {NaC$_6$H$^+$}    & 4.75 10$^{-12}$   &  $-$0.972 \\
       {AlC$_6$H$^+$}    & 4.30 10$^{-11}$   &  $-$0.952 \\
       {MgC$_6$H$^+$}    & 5.19 10$^{-10}$   &  $-$0.706 \\
       {MgC$_5$N$^+$}    & 1.06 10$^{-9}$    &  $-$0.584 \\
       {MgC$_4$H$^+$}    & 7.22 10$^{-12}$   &  $-$0.936 \\
       {MgC$_3$N$^+$}    & 2.02 10$^{-14}$   &  $-$0.878 \\
        \hline
    \end{tabular}
\end{table}

\section{Chemical model of IRC\,+10216: astrophysical implications}

Metal-bearing carbon chains have been so far observed only in circumstellar envelopes around carbon-rich AGB stars, and most of them exclusively in one object, IRC\,+10216. To evaluate the impact of the radiative association rate coefficients calculated here on the abundances of metal-bearing molecules in IRC\,+10216 we implemented them in a chemical model. The formation scenario is based on the original idea of \cite{Petrie1996}. The model is based on that presented in \cite{Agundez2017}, where the chemical network was expanded to include metal chemistry \citep{Cabezas2013,Cabezas2023,Cernicharo2019,Cernicharo2023,Pardo2021}. Rate coefficients for reactions of radiative association between metal ionized atoms and carbon chains were taken from theoretical calculations, in the case of closed electronic shell species (polyynes and cyanopolyynes) from \cite{Dunbar-Petrie:02} and \cite{Petrie2004}, and in the case of carbon chain radicals from this study (see Table\,\ref{tab:RArates}). For reactions involving Al$^+$ and Na$^+$ whose rate coefficients have not been explicitly calculated here, we adopted the reaction Mg$^+$ + C$_6$H as reference and scaled the rate coefficient as follows
\begin{equation}
\rm \emph{k} [M^+ + R] = \emph{k} [Mg^+ + R] \frac{\emph{k} [M^+ + C_6H]}{\emph{k} [Mg^+ + C_6H]},
\end{equation}
where M$^+$ stands for Al$^+$ or Na$^+$ and R stands for the carbon chain radical C$_4$H, C$_3$N, or C$_5$N. For larger radicals we adopted the rate coefficient of either C$_6$H or C$_5$N, depending on the type of carbon chain. The metal cationic complexes MR$^+$ were assumed to be destroyed by dissociative recombination with electrons
{,  and here we employ the same used before}
  , where the branching ratios of the different fragments were tuned to reproduce observed abundance ratios in IRC\,+10216, as in previous studies \citep{Cabezas2013,Cernicharo2019,Pardo2021}, and by reaction with H atoms. The metal-carbon neutral fragments formed in the dissociative recombination of MR$^+$ were assumed to be destroyed by reactions with electrons and H atoms, and by UV photodissociation. The initial abundances of Mg, Al, and Na in the gas phase (relative to H$_2$) were chosen as 3.6\,$\times$\,10$^{-6}$, 7.4\,$\times$\,10$^{-9}$, 4.2\,$\times$\,10$^{-7}$, respectively, to reproduce the column densities of MgNC, AlNC, and NaC$_3$N, respectively, observed in IRC\,+10216. In this comparison exercise between calculated and observed column densities, we adopted twice the radial column density computed with the chemical model.

\begin{figure}
\resizebox{\hsize}{!}{\includegraphics[width=\linewidth]{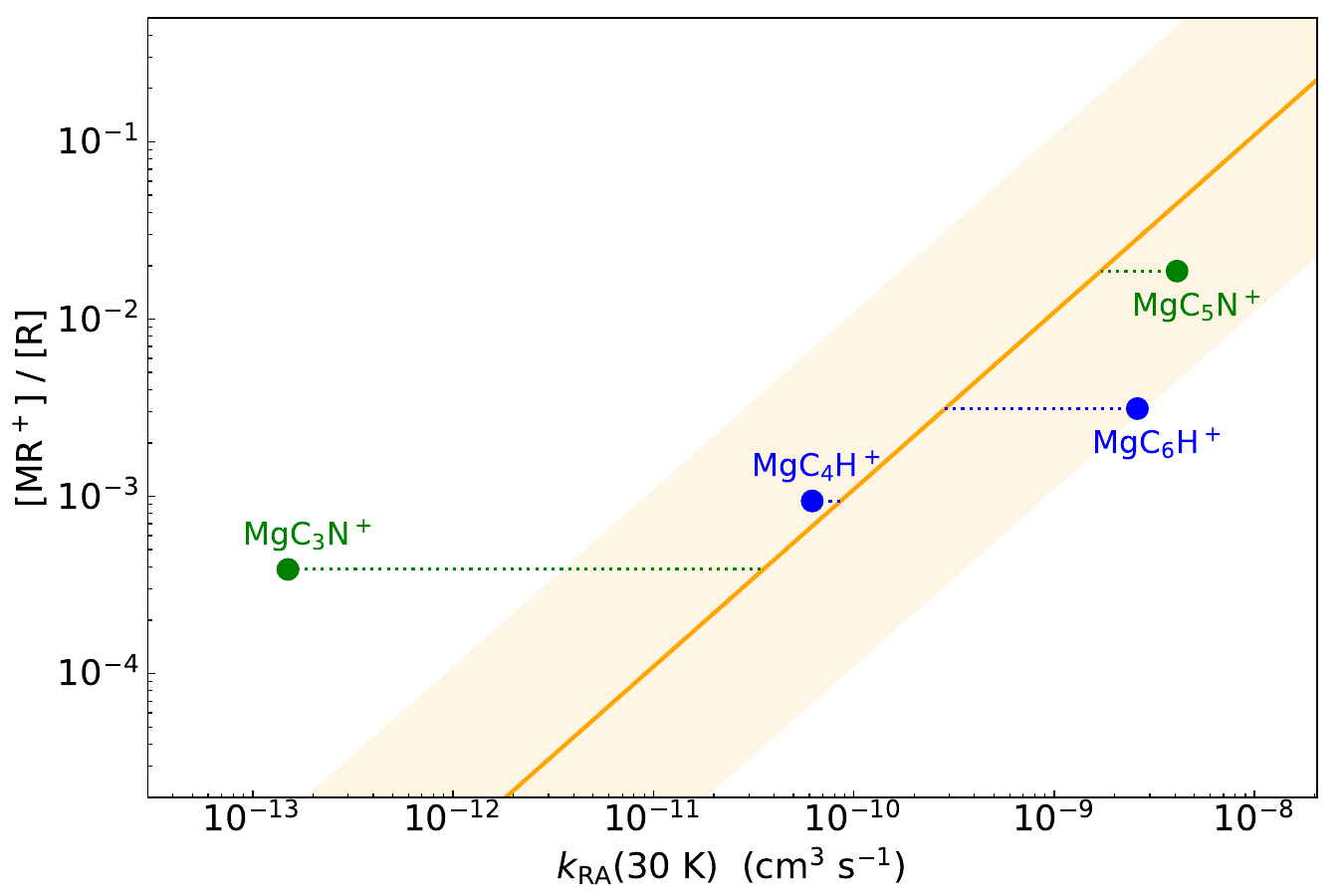}}
\caption{Abundance ratios [MR$^+$]/[R] observed in IRC\,+10216 as a function of the rate coefficient of radiative association between M$^+$ and R calculated here at 30 K. The shadowed region shows the expected [MR$^+$]/[R] values according to the theoretical relationship given in Eq.\,(\ref{eq:mr+}), with an uncertainty of a factor of ten. Abundance ratios come from the column densities observed: $N$(MgC$_4$H$^+$)\,=\,4.8\,$\times$\,10$^{11}$ cm$^{-2}$, $N$(MgC$_6$H$^+$)\,=\,2.5\,$\times$\,10$^{11}$ cm$^{-2}$, $N$(MgC$_3$N$^+$)\,=\,1.2\,$\times$\,10$^{11}$ cm$^{-2}$, and $N$(MgC$_5$N$^+$)\,=\,1.1\,$\times$\,10$^{11}$ cm$^{-2}$ \citep{Cernicharo2023}, $N$(C$_4$H)\,=\,5.1\,$\times$\,10$^{14}$ cm$^{-2}$ \citep{Oyama2020}, $N$(C$_6$H)\,=\,8.0\,$\times$\,10$^{13}$ cm$^{-2}$ \citep{Cernicharo2008}, $N$(C$_3$N)\,=\,3.1\,$\times$\,10$^{14}$ cm$^{-2}$ \citep{Thaddeus2008}, and $N$(C$_5$N)\,=\,5.9\,$\times$\,10$^{12}$ cm$^{-2}$ \citep{Guelin1998}.}
\label{fig:cations}
\end{figure}

Let us consider a simplified chemical scheme in which a given metal cationic complex MR$^+$ is formed by the corresponding radiative association and it is mostly destroyed by dissociative recombination with electrons, which in fact dominates over the reaction with H atoms according to the chemical model. If we equal the formation and destruction rates we arrive at
\begin{equation}
\rm \frac{[MR^+]}{[R]} = \frac{[M^+]}{\emph{k}_{DR} ~ [e^-]} ~ \emph{k}_{RA}, \label{eq:mr+}
\end{equation}
where the brackets refer to fractional abundances relative to H$_2$, and $k_{\rm RA}$ and $k_{\rm DR}$ stand for the rate coefficients of the radiative association M$^+$ + R and the dissociative recombination MR$^+$ + e$^-$, respectively. If we focus on metal cationic complexes of magnesium in the region of the outer envelope of IRC\,+10216 where these species are abundant, to a first approximation [M$^+$] and [e$^-$] can be taken as constant and $k_{\rm DR}$ can be assumed to take the same value regardless of the cation MR$^+$. Therefore, according to Eq.\,(\ref{eq:mr+}), the abundance ratio [MR$^+$]/[R] is expected to be directly proportional to the corresponding rate coefficient $k_{\rm RA}$. The constant of proportionality can be estimated adopting [Mg$^+$]\,$\sim$\,7.0\,$\times$\,10$^{-8}$ and [e$^-$]\,$\sim$\,2.0\,$\times$\,10$^{-8}$ from the chemical model, and evaluating the $k_{\rm DR}$ at 30 K (the temperature in the region where MR$^+$ are expected; \citealt{Guelin2018}) from the canonical expression 10$^{-7}$ cm$^3$ s$^{-1}$ ($T$/300)$^{-0.5}$. Figure\,\ref{fig:cations} shows the observed abundance ratios [MR$^+$]/[R] versus the calculated rate coefficient of radiative association M$^+$ + R. It is seen that MgC$_4$H$^+$, MgC$_6$H$^+$, and MgC$_5$N$^+$ agree with the theoretical linear relationship within one order of magnitude, while MgC$_3$N$^+$ deviates significantly, most likely because the calculated rate coefficient for the radiative association Mg$^+$ + C$_3$N is too low by about two orders of magnitude.

\begin{figure*}
\centering
\includegraphics[width=17cm]{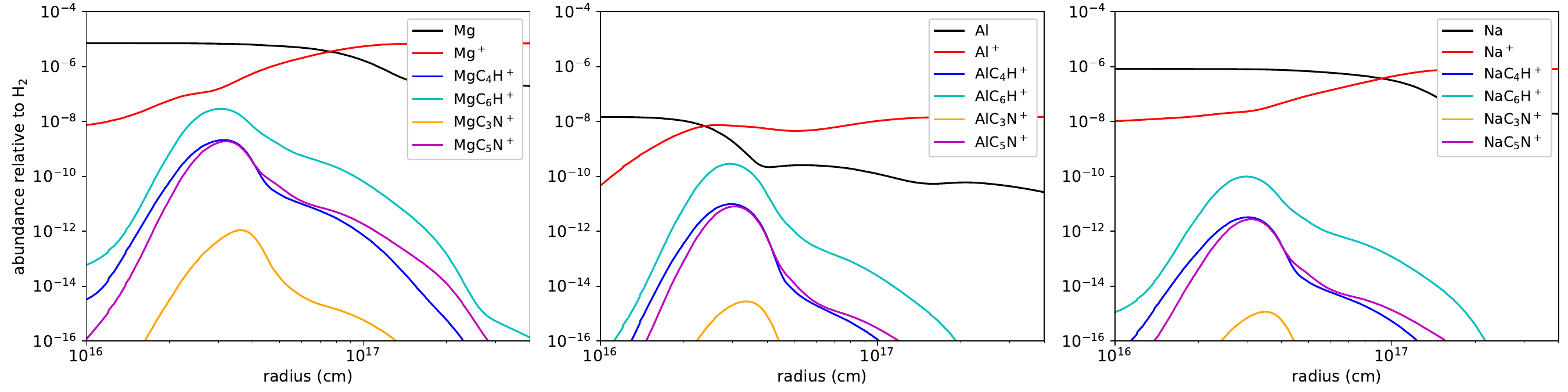}
\caption{Abundances for Mg-, Al-, and Na-bearing cations calculated with the chemical model of IRC\,+10216.}
\label{fig:model}
\end{figure*}

In Fig.\,\ref{fig:model} we show the calculated abundances of the metal cationic complexes as a function of radial distance to the star. Most of the observational constraints are available for magnesium, in which case the four cations MgC$_4$H$^+$, MgC$_6$H$^+$, MgC$_3$N$^+$, and MgC$_5$N$^+$ have been detected in IRC\,+10216 \citep{Cernicharo2023}. The conclusions from the chemical model regarding magnesium are similar to those obtained before with the simplified chemical scheme. The calculated abundances of MgC$_4$H$^+$, MgC$_6$H$^+$, and MgC$_5$N$^+$ are consistent with observations within one order of magnitude. However, the chemical model underestimates the abundance of MgC$_3$N$^+$ by a factor of 300-400, in line with the conclusion extracted before. Unless MgC$_3$N$^+$ can be efficiently formed by alternative routes to the radiative association between Mg$^+$ and C$_3$N, the immediate conclusion is that the calculated rate coefficient for this latter reaction is too low by about two orders of magnitude. The situation resembles that found regarding the formation of negative ions, where calculated rate coefficients for radiative electron attachment to neutral radicals are consistent with observations for large species but have troubles for medium size species, such as C$_4$H and C$_3$N \citep{Petrie1997,Herbst2008,Agundez2023}. The metal-carbon cations involving Al and Na are calculated to be around two orders of magnitude less abundant than those of Mg, because the analogous radiative association reactions are slower and the metal has a lower gas-phase abundance. According to this, it seems unlikely that Na- or Al- cations of the type NaC$_6$H$^+$ or AlC$_6$H$^+$ are detected in IRC\,+10216.

\section{Conclusions}

The formation of $\text{MC}_6\text{H}^+$ adducts via radiative association is investigated for $\text{M} = \text{Mg, Al, and Na}$ using statistical methods based on high-level ab initio calculations. The results indicate that binding energies decrease in the order $\text{Mg}^+ > \text{Al}^+ > \text{Na}^+$, a trend directly proportional to the ionization energy of the metal cations. For the $\text{Mg}^+$ and $\text{Al}^+$ adducts, an electronic crossing occurs with an excited state—characterized as $\text{M}^{2+}\text{C}_6\text{H}^-$—resulting in a $\Sigma$ electronic ground state for the linear equilibrium configuration. Conversely, the high ionization energy of $\text{Na}^+$ precludes this crossing and, consequently, the more weakly bound $\text{NaC}_6\text{H}^+$ linear adduct retains the $\Pi$ symmetry of the ground state of C$_6$H.

The reorientation of the two precursors along the capture  is investigated using a quasi-classical method \citep{Su-Bowers:73,Su-Bowers:75}, allowing to determine a locking parameter producing a capture reduction of 3/4 with respect to the locked dipole monodimensional model. Also, low vibrational frequencies are limited to a minimum value of 100 cm$^{-1}$, because they have larger relative errors and can be considered to be decoupled from the high frequency vibrational model. These two considerations yield to a reduction of the radiative association rate coefficients reported previously \citep{Cernicharo2023}. It is found that the rate coefficient shows a progression with the accompanying cation proportional to the binding energies, $\text{Mg}^+ > \text{Al}^+ > \text{Na}^+$. 

The new rate coefficients are used in a chemical model of the carbon-rich envelope IRC\,+1216, where these Mg-bearing carbon cations have been observed. The abundances of $\text{MgC}_4\text{H}^+$, MgC$_5$N$^+$ and $\text{MgC}_6\text{H}^+$ are within one order of magnitude consistent with observations, corroborating that radiative association is a plausible formation mechanism for these adducts, as first proposed by \cite{Petrie1996}. However, the abundance of MgC$_3$N$^+$ in the model is about two orders of magnitude smaller than the observed one, which indicates that either the calculated rate coefficient is too low or other formation or destruction mechanisms operate for this molecule. 

Interestingly, the abundances of Al and Na adducts, are smaller than the corresponding Mg ones by about two orders of magnitude. The reason is the lower radiative association rate (due to lower binding energies) and that the parent metal ions are also less abundant. As a consequence, we conclude that it would be very difficult to detect Na- and Al-bearing carbon cations in IRC\,+10216.

\begin{acknowledgements}

The research leading to these results has received funding from Ministerio de Ciencia, Innovación y Universidades (Spain), under grants PID2021-122549NB-C21, PID2023-147545NB-I00, and PID2024-156686NB-I00. GM acknowledges support by ERC grant No. 101218790 (Isocosmos) funded by the European Union. Views and opinions expressed are however
those of the author(s) only and do not necessarily reflect those of the European Union or the European Research Council Executive Agency. Neither the European Union nor the granting authority can be held responsible for them. GM also acknowledges the support of the grant RYC2022-035442-I funded by MICIU/AEI/10.13039/501100011033 and ESF+ and project 20245AT016
(Proyectos Intramurales CSIC). NB acknowledge IDUB program for coninancing the project: Nobelium Joining Gda\'nsk Tech Research community, contract number: DEC1/12025/IDUB/I.1a/038122. PSZ is  grateful for the support of the National Science Centre, Poland (NCN) through grant ``Sonata Bis 9'' No. 2019/34/E/ST4/00407. Computational assistance was provided by the Supercomputer facilities of Lusitania founded by the C\'enitS and Computaex Foundation.

\end{acknowledgements}

\bibliographystyle{aa}

\begin{thebibliography}{83}
\expandafter\ifx\csname natexlab\endcsname\relax\def\natexlab#1{#1}\fi

\bibitem[{{Ag{\'u}ndez} {et~al.}(2014){Ag{\'u}ndez}, {Cernicharo}, \&
  {Gu{\'e}lin}}]{Agundez2014}
{Ag{\'u}ndez}, M., {Cernicharo}, J., \& {Gu{\'e}lin}, M. 2014, \aap, 570, A45

\bibitem[{{Ag{\'u}ndez} {et~al.}(2017){Ag{\'u}ndez}, {Cernicharo},
  {Quintana-Lacaci}, {Castro-Carrizo}, {Velilla Prieto}, {Marcelino},
  {Gu{\'e}lin}, {Joblin}, {Mart{\'\i}n-Gago}, {Gottlieb}, {Patel}, \&
  {McCarthy}}]{Agundez2017}
{Ag{\'u}ndez}, M., {Cernicharo}, J., {Quintana-Lacaci}, G., {et~al.} 2017,
  \aap, 601, A4

\bibitem[{{Ag{\'u}ndez} {et~al.}(2023){Ag{\'u}ndez}, {Marcelino}, {Tercero},
  {Jim{\'e}nez-Serra}, \& {Cernicharo}}]{Agundez2023}
{Ag{\'u}ndez}, M., {Marcelino}, N., {Tercero}, B., {Jim{\'e}nez-Serra}, I., \&
  {Cernicharo}, J. 2023, \aap, 677, A106

\bibitem[{Basilio {et~al.}(2026)Basilio, Molpeceres, Wolff, Pereira-Santaella,
  \& Oliveira}]{basilio_double_2026}
Basilio, J. H.~C., Molpeceres, G., Wolff, W., Pereira-Santaella, M., \&
  Oliveira, R.~R. 2026, ACS Earth and Space Chemistry, 10, 370

\bibitem[{{Cabezas} {et~al.}(2013){Cabezas}, {Cernicharo}, {Alonso},
  {Ag{\'u}ndez}, {Mata}, {Gu{\'e}lin}, \& {Pe{\~n}a}}]{Cabezas2013}
{Cabezas}, C., {Cernicharo}, J., {Alonso}, J.~L., {et~al.} 2013, \apj, 775, 133

\bibitem[{{Cabezas} {et~al.}(2023){Cabezas}, {Pardo}, {Ag{\'u}ndez}, {Tercero},
  {Marcelino}, {Endo}, {de Vicente}, {Gu{\'e}lin}, \&
  {Cernicharo}}]{Cabezas2023}
{Cabezas}, C., {Pardo}, J.~R., {Ag{\'u}ndez}, M., {et~al.} 2023, \aap, 672, L12

\bibitem[{{Cernicharo} {et~al.}(2019{\natexlab{a}}){Cernicharo}, {Cabezas},
  {Pardo}, {Ag{\'u}ndez}, {Berm{\'u}dez}, {Velilla-Prieto}, {Tercero},
  {L{\'o}pez-P{\'e}rez}, {Gallego}, {Fonfr{\'\i}a}, {Quintana-Lacaci},
  {Gu{\'e}lin}, \& {Endo}}]{Cernicharo2019}
{Cernicharo}, J., {Cabezas}, C., {Pardo}, J.~R., {et~al.} 2019{\natexlab{a}},
  \aap, 630, L2

\bibitem[{{Cernicharo} {et~al.}(2023){Cernicharo}, {Cabezas}, {Pardo},
  {Ag{\'u}ndez}, {Roncero}, {Tercero}, {Marcelino}, {Gu{\'e}lin}, {Endo}, \&
  {de Vicente}}]{Cernicharo2023}
{Cernicharo}, J., {Cabezas}, C., {Pardo}, J.~R., {et~al.} 2023, \aap, 672, L13

\bibitem[{Cernicharo \& Gu\'elin(1987)}]{Cernicharo-Guelin:87}
Cernicharo, J. \& Gu\'elin, M. 1987, Astron. AstroPhys., 183, L10

\bibitem[{{Cernicharo} {et~al.}(2008){Cernicharo}, {Gu{\'e}lin}, {Ag{\'u}ndez},
  {McCarthy}, \& {Thaddeus}}]{Cernicharo2008}
{Cernicharo}, J., {Gu{\'e}lin}, M., {Ag{\'u}ndez}, M., {McCarthy}, M.~C., \&
  {Thaddeus}, P. 2008, \apjl, 688, L83

\bibitem[{{Cernicharo} {et~al.}(2019{\natexlab{b}}){Cernicharo},
  {Velilla-Prieto}, {Ag{\'u}ndez}, {Pardo}, {Fonfr{\'\i}a}, {Quintana-Lacaci},
  {Cabezas}, {Berm{\'u}dez}, \& {Gu{\'e}lin}}]{Cernicharo2019_canc}
{Cernicharo}, J., {Velilla-Prieto}, L., {Ag{\'u}ndez}, M., {et~al.}
  2019{\natexlab{b}}, \aap, 627, L4

\bibitem[{{Changala} {et~al.}(2022){Changala}, {Gupta}, {Cernicharo}, {Pardo},
  {Ag{\'u}ndez}, {Cabezas}, {Tercero}, {Gu{\'e}lin}, \&
  {McCarthy}}]{Changala2022}
{Changala}, P.~B., {Gupta}, H., {Cernicharo}, J., {et~al.} 2022, \apjl, 940,
  L42

\bibitem[{Clary(1985)}]{Clary:85}
Clary, D.~C. 1985, Mol. Phys., 54, 605

\bibitem[{{Cordiner} \& {Millar}(2009)}]{Cordiner2009}
{Cordiner}, M.~A. \& {Millar}, T.~J. 2009, \apj, 697, 68

\bibitem[{Dunbar \& Petrie(2002)}]{Dunbar-Petrie:02}
Dunbar, R.~C. \& Petrie, S. 2002, AstroPhys. J., 564, 792

\bibitem[{Dunning \& Jr.(1989)}]{Dunning:89}
Dunning, T.~H. \& Jr. 1989, J. Chem. Phys., 90, 1007

\bibitem[{Garand {et~al.}(2010)Garand, Yacovitch, Zhou, Sheehan, \&
  Numark}]{Garand-etal:10}
Garand, E., Yacovitch, T.~I., Zhou, J., Sheehan, S.~M., \& Numark, D.~M. 2010,
  Chem. Sci., 1, 192

\bibitem[{{Ginsburg} {et~al.}(2019){Ginsburg}, {McGuire}, {Plambeck}, {Bally},
  {Goddi}, \& {Wright}}]{Ginsburg2019}
{Ginsburg}, A., {McGuire}, B., {Plambeck}, R., {et~al.} 2019, \apj, 872, 54

\bibitem[{{Gu\'elin} {et~al.}(1986){Gu\'elin}, {Cernicharo}, {Kahane}, \&
  {Gomez-Gonzales}}]{Guelin1986}
{Gu\'elin}, M., {Cernicharo}, J., {Kahane}, C., \& {Gomez-Gonzales}, J. 1986,
  \aap, 157, L17

\bibitem[{{Guelin} {et~al.}(1993){Guelin}, {Lucas}, \&
  {Cernicharo}}]{Guelin1993}
{Guelin}, M., {Lucas}, R., \& {Cernicharo}, J. 1993, \aap, 280, L19

\bibitem[{{Guelin} {et~al.}(1998){Guelin}, {Neininger}, \&
  {Cernicharo}}]{Guelin1998}
{Guelin}, M., {Neininger}, N., \& {Cernicharo}, J. 1998, \aap, 335, L1

\bibitem[{{Gu{\'e}lin} {et~al.}(2018){Gu{\'e}lin}, {Patel}, {Bremer},
  {Cernicharo}, {Castro-Carrizo}, {Pety}, {Fonfr{\'\i}a}, {Ag{\'u}ndez},
  {Santander-Garc{\'\i}a}, {Quintana-Lacaci}, {Velilla Prieto}, {Blundell}, \&
  {Thaddeus}}]{Guelin2018}
{Gu{\'e}lin}, M., {Patel}, N.~A., {Bremer}, M., {et~al.} 2018, \aap, 610, A4

\bibitem[{Guo {et~al.}(2018)Guo, Riplinger, Becker, Liakos, Minenkov, Cavallo,
  \& Neese}]{guo_communication_2018}
Guo, Y., Riplinger, C., Becker, U., {et~al.} 2018, The Journal of Chemical
  Physics, 148, 011101

\bibitem[{{Gupta} {et~al.}(2024){Gupta}, {Changala}, {Cernicharo}, {Pardo},
  {Ag{\'u}ndez}, {Cabezas}, {Tercero}, {Gu{\'e}lin}, \& {McCarthy}}]{Gupta2024}
{Gupta}, H., {Changala}, P.~B., {Cernicharo}, J., {et~al.} 2024, \apjl, 966,
  L28

\bibitem[{Hapka {et~al.}(2012)Hapka, Zuchowski, \& anf
  G.~Chalasiński}]{Hapka-etal:12}
Hapka, M., Zuchowski, P.~S., \& anf G.~Chalasiński, M. M.~S. 2012, J. Chem.
  Phys., 137, 164104

\bibitem[{Herbst(1976)}]{Herbst:76}
Herbst, E. 1976, AstroPhys. J., 205, 94

\bibitem[{Herbst(1982)}]{Herbst:82}
Herbst, E. 1982, Chem. Phys., 65, 185

\bibitem[{Herbst(1987)}]{Herbst:87}
Herbst, E. 1987, AstroPhys. J., 313, 867

\bibitem[{{Herbst} \& {Osamura}(2008)}]{Herbst2008}
{Herbst}, E. \& {Osamura}, Y. 2008, \apj, 679, 1670

\bibitem[{{I. S. Gradshteyn and I. M. Ryzhik}(1980)}]{Gradshteyn-Ryzhik:80}
{I. S. Gradshteyn and I. M. Ryzhik}. 1980, Table of Integrals, Series and
  Products (Academic Press)

\bibitem[{{Kami{\'n}ski} {et~al.}(2013){Kami{\'n}ski}, {Gottlieb}, {Menten},
  {Patel}, {Young}, {Br{\"u}nken}, {M{\"u}ller}, {McCarthy}, {Winters}, \&
  {Decin}}]{Kaminski2013}
{Kami{\'n}ski}, T., {Gottlieb}, C.~A., {Menten}, K.~M., {et~al.} 2013, \aap,
  551, A113

\bibitem[{{Kawaguchi} {et~al.}(1993){Kawaguchi}, {Kagi}, {Hirano}, {Takano}, \&
  {Saito}}]{Kawaguchi1993}
{Kawaguchi}, K., {Kagi}, E., {Hirano}, T., {Takano}, S., \& {Saito}, S. 1993,
  \apjl, 406, L39

\bibitem[{Kirby \& Dalgarno(1978)}]{Kirby-Dalgarno:78}
Kirby, K. \& Dalgarno, A. 1978, AstroPhys. J., 224, 444

\bibitem[{{Koelemay} \& {Ziurys}(2023)}]{Koelemay2023}
{Koelemay}, L.~A. \& {Ziurys}, L.~M. 2023, \apjl, 958, L6

\bibitem[{Kramida {et~al.}(2024)Kramida, {Yu.~Ralchenko}, Reader, \& {and NIST
  ASD Team}}]{NIST_AtomicSpectraDatabase}
Kramida, A., {Yu.~Ralchenko}, Reader, J., \& {and NIST ASD Team}. 2024, {NIST
  Atomic Spectra Database (ver. 5.12), [Online]. Available:
  {\tt{https://physics.nist.gov/asd}} [2026, April 21]. National Institute of
  Standards and Technology, Gaithersburg, MD.}

\bibitem[{{L. M. Bass and P. R. Kemper and V. G. Anicich and M. T.
  Bowers}(1981)}]{Bass-etal:81}
{L. M. Bass and P. R. Kemper and V. G. Anicich and M. T. Bowers}. 1981, J. Am.
  Chem. Soc., 103, 5283

\bibitem[{Levine \& Bernstein(1987)}]{Levine-Bernstein:87}
Levine, R.~D. \& Bernstein, R.~B. 1987, Molecular Reaction Dynamics and
  Chemical Reactivity (Oxford: Oxford University Press)

\bibitem[{Maeda {et~al.}(2023{\natexlab{a}})Maeda, Harabuchi, Hayashi, \&
  Mita}]{maeda_toward_2023}
Maeda, S., Harabuchi, Y., Hayashi, H., \& Mita, T. 2023{\natexlab{a}}, Annual
  Review of Physical Chemistry, 74, 287

\bibitem[{Maeda {et~al.}(2023{\natexlab{b}})Maeda, Harabuchi, Sumiya,
  {et~al.}}]{grrm23}
Maeda, S., Harabuchi, Y., Sumiya, Y., {et~al.} 2023{\natexlab{b}}, GRRM23, see
  \url{https://global.hpc.co.jp/products/grrm23/}

\bibitem[{Maeda {et~al.}(2018)Maeda, Harabuchi, Takagi, Saita, Suzuki, Ichino,
  Sumiya, Sugiyama, \& Ono}]{maeda_implementation_2018}
Maeda, S., Harabuchi, Y., Takagi, M., {et~al.} 2018, Journal of Computational
  Chemistry, 39, 233

\bibitem[{Maeda {et~al.}(2013)Maeda, Ohno, \& Morokuma}]{maeda_systematic_2013}
Maeda, S., Ohno, K., \& Morokuma, K. 2013, Physical Chemistry Chemical Physics,
  15, 3683

\bibitem[{Mauron \& Huggins(2010)}]{Mauron-Huggins:10}
Mauron, N. \& Huggins, P.~J. 2010, Astron. Astrophys., 513, A31

\bibitem[{{Millar}(2008)}]{Millar2008}
{Millar}, T.~J. 2008, \apss, 313, 223

\bibitem[{{Millar} {et~al.}(2024){Millar}, {Walsh}, {Van de Sande}, \&
  {Markwick}}]{Millar2024}
{Millar}, T.~J., {Walsh}, C., {Van de Sande}, M., \& {Markwick}, A.~J. 2024,
  \aap, 682, A109

\bibitem[{Miller(1987)}]{Miller:87}
Miller, W.~H. 1987, Chem. Rev., 87, 19

\bibitem[{Moran \& Hamill(1963)}]{Moran-Hamill:63}
Moran, T.~F. \& Hamill, W.~H. 1963, J. Chem. Phys., 39, 1413

\bibitem[{Mulliken(1955)}]{Mulliken:55}
Mulliken, R.~S. 1955, J. Chem. Phys., 23, 1833

\bibitem[{{Oyama} {et~al.}(2020){Oyama}, {Ozaki}, {Sumiyoshi}, {Araki},
  {Takano}, {Kuze}, \& {Tsukiyama}}]{Oyama2020}
{Oyama}, T., {Ozaki}, H., {Sumiyoshi}, Y., {et~al.} 2020, \apj, 890, 39

\bibitem[{{Pardo} {et~al.}(2021){Pardo}, {Cabezas}, {Fonfr{\'\i}a},
  {Ag{\'u}ndez}, {Tercero}, {de Vicente}, {Gu{\'e}lin}, \&
  {Cernicharo}}]{Pardo2021}
{Pardo}, J.~R., {Cabezas}, C., {Fonfr{\'\i}a}, J.~P., {et~al.} 2021, \aap, 652,
  L13

\bibitem[{Perdew {et~al.}(2009)Perdew, Ruzsinszky, Csonka, Constantin, \&
  Sun}]{perdew_workhorse_2009}
Perdew, J.~P., Ruzsinszky, A., Csonka, G.~I., Constantin, L.~A., \& Sun, J.
  2009, Physical Review Letters, 103, 026403

\bibitem[{{Petrie}(1996)}]{Petrie1996}
{Petrie}, S. 1996, \mnras, 282, 807

\bibitem[{Petrie(2004)}]{Petrie2004}
Petrie, S. 2004, Australian Journal of Chemistry, 57, 67

\bibitem[{{Petrie} \& {Herbst}(1997)}]{Petrie1997}
{Petrie}, S. \& {Herbst}, E. 1997, \apj, 491, 210

\bibitem[{Quack \& Troe(1974)}]{Quack-Troe:74}
Quack, M. \& Troe, J. 1974, Ber. Bunsenges. Phys. Chem, 78, 240

\bibitem[{Rappoport \& Furche(2010)}]{rappoport_property-optimized_2010}
Rappoport, D. \& Furche, F. 2010, The Journal of Chemical Physics, 133, 134105

\bibitem[{Ribeiro {et~al.}(2011)Ribeiro, Marenich, Cramer, \&
  Truhlar}]{Ribeiro-etal:11}
Ribeiro, R.~F., Marenich, A.~V., Cramer, C.~J., \& Truhlar, D.~G. 2011, J.
  Phys. Chem. B, 115, 14556

\bibitem[{{S. J. Klippenstein and Y.C. Yang and V. Ryzhov and R.
  Dunbar}(1996)}]{Klippenstein-etal:96}
{S. J. Klippenstein and Y.C. Yang and V. Ryzhov and R. Dunbar}. 1996, J. Chem.
  Phys., 104, 4502

\bibitem[{Sanz-Sanz {et~al.}(2015)Sanz-Sanz, Aguado, Roncero, \&
  Naumkin}]{Sanz-Sanz-etal:15}
Sanz-Sanz, C., Aguado, A., Roncero, O., \& Naumkin, F. 2015, J. Chem. Phys.,
  143, 234303

\bibitem[{Smith {et~al.}(1983)Smith, Adams, Alge, \& Herbst}]{Smith-etal:83}
Smith, D., Adams, N., Alge, E., \& Herbst, E. 1983, AstroPhys. J., 272, 365

\bibitem[{Stein \& Rabinovitch(1973)}]{Stein-Rabinovitch:73}
Stein, S.~E. \& Rabinovitch, B.~S. 1973, J. Chem. Phys., 58, 2438

\bibitem[{Su \& Bowers(1973)}]{Su-Bowers:73}
Su, T. \& Bowers, M.~T. 1973, J. Chem. Phys., 58, 3027

\bibitem[{Su \& Bowers(1975)}]{Su-Bowers:75}
Su, T. \& Bowers, M.~T. 1975, J. Chem. Phys., 63

\bibitem[{{T. SU and W. J. Chesnavich}(1982)}]{Su-Chesnavich:82}
{T. SU and W. J. Chesnavich}. 1982, J. Chem. Phys., 76, 5183

\bibitem[{{Tachibana} {et~al.}(2019){Tachibana}, {Kamizuka}, {Hirota}, {Sakai},
  {Oya}, {Takigawa}, \& {Yamamoto}}]{Tachibana2019}
{Tachibana}, S., {Kamizuka}, T., {Hirota}, T., {et~al.} 2019, \apjl, 875, L29

\bibitem[{Tao {et~al.}(2003)Tao, Perdew, Staroverov, \&
  Scuseria}]{tao_climbing_2003}
Tao, J., Perdew, J.~P., Staroverov, V.~N., \& Scuseria, G.~E. 2003, Physical
  Review Letters, 91, 146401

\bibitem[{Taylor {et~al.}(1998)Taylor, Xu, \& Neumark}]{Taylor-etal:98}
Taylor, Xu, C., \& Neumark, D.~M. 1998, J. Chem. Phys., 108, 10018

\bibitem[{{Tenenbaum} \& {Ziurys}(2009)}]{Tenenbaum2009}
{Tenenbaum}, E.~D. \& {Ziurys}, L.~M. 2009, \apjl, 694, L59

\bibitem[{{Tenenbaum} \& {Ziurys}(2010)}]{Tenenbaum2010}
{Tenenbaum}, E.~D. \& {Ziurys}, L.~M. 2010, \apjl, 712, L93

\bibitem[{{Thaddeus} {et~al.}(2008){Thaddeus}, {Gottlieb}, {Gupta},
  {Br{\"u}nken}, {McCarthy}, {Ag{\'u}ndez}, {Gu{\'e}lin}, \&
  {Cernicharo}}]{Thaddeus2008}
{Thaddeus}, P., {Gottlieb}, C.~A., {Gupta}, H., {et~al.} 2008, \apj, 677, 1132

\bibitem[{{Turner}(1991)}]{Turner1991}
{Turner}, B.~E. 1991, \apj, 376, 573

\bibitem[{{Turner} {et~al.}(2005){Turner}, {Petrie}, {Dunbar}, \&
  {Langston}}]{Turner2005}
{Turner}, B.~E., {Petrie}, S., {Dunbar}, R.~C., \& {Langston}, G. 2005, \apj,
  621, 817

\bibitem[{Velmiskina {et~al.}(2025)Velmiskina, Malyshev, Gerasimov, \&
  Medvedev}]{Velmiskina-etal:25}
Velmiskina, J.~A., Malyshev, V.~I., Gerasimov, I.~S., \& Medvedev, M.~G. 2025,
  J. Chem. Phys., 163, 124115

\bibitem[{{W. J. Chesnavich and T. Su and M. T.
  Bowers}(1980)}]{Chesnavich-etal:80}
{W. J. Chesnavich and T. Su and M. T. Bowers}. 1980, J. Chem. Phys., 72, 2641

\bibitem[{Weigend \& Ahlrichs(2005)}]{Weigend2005}
Weigend, F. \& Ahlrichs, R. 2005, Physical Chemistry Chemical Physics, 7, 3297

\bibitem[{Werner \& Knowles(1988{\natexlab{a}})}]{Werner-Knowles:88}
Werner, H.~J. \& Knowles, P.~J. 1988{\natexlab{a}}, J. Chem. Phys., 89, 5803

\bibitem[{Werner \& Knowles(1988{\natexlab{b}})}]{Werner-Knowles:88b}
Werner, H.~J. \& Knowles, P.~J. 1988{\natexlab{b}}, Chem. Phys. Lett., 145, 514

\bibitem[{Werner {et~al.}(2012)Werner, Knowles, Knizia, Manby, \&
  Sch{\"u}tz}]{MOLPRO-WIREs}
Werner, H.-J., Knowles, P.~J., Knizia, G., Manby, F.~R., \& Sch{\"u}tz, M.
  2012, WIREs Comput Mol Sci, 2, 242

\bibitem[{Whiting \& Nicholls(1974)}]{Whiting-Nicholls:74}
Whiting, E.~E. \& Nicholls, R.~W. 1974, Astrophys. J. Supp. Series, 27, 1

\bibitem[{Whiting {et~al.}(1980)Whiting, Schadee, Taum, Hougen, \&
  Nicholls}]{Whiting-etal:80}
Whiting, E.~E., Schadee, A., Taum, J.~B., Hougen, J.~T., \& Nicholls, R.~W.
  1980, J. Mol. Spect., 80, 249

\bibitem[{Woon \& Herbst(2009)}]{Woon-Herbst:09}
Woon, D.~E. \& Herbst, E. 2009, ApJ. Sup. Series, 185, 273

\bibitem[{{Zack} {et~al.}(2011){Zack}, {Halfen}, \& {Ziurys}}]{Zack2011}
{Zack}, L.~N., {Halfen}, D.~T., \& {Ziurys}, L.~M. 2011, \apjl, 733, L36

\bibitem[{{Ziurys} {et~al.}(1995){Ziurys}, {Apponi}, {Guelin}, \&
  {Cernicharo}}]{Ziurys1995}
{Ziurys}, L.~M., {Apponi}, A.~J., {Guelin}, M., \& {Cernicharo}, J. 1995,
  \apjl, 445, L47

\bibitem[{{Ziurys} {et~al.}(2002){Ziurys}, {Savage}, {Highberger}, {Apponi},
  {Gu{\'e}lin}, \& {Cernicharo}}]{Ziurys2002}
{Ziurys}, L.~M., {Savage}, C., {Highberger}, J.~L., {et~al.} 2002, \apjl, 564,
  L45

\end{thebibliography}

\clearpage

\begin{appendix}

\section{Quasi-classical capture rates \label{appendix:qct}}

The quasi-classical capture rates for Mg$^+$+C$_6$H are calculated considering a rigid C$_6$H as a pseudo diatomic molecule, (C$_3$)-(C$_3$H), with and equilibrium distance fitted to reproduce the C$_6$H rotational constant. Two model potentials were used to describe the long-range interactions, $V_1$, given in Eq.\,(\ref{eq:ion-dipole-interaction}), and $V_2$, to add the charge-induce-dipole interaction and defined as
\begin{eqnarray}\label{eq:charge-induced-dipole}
V_2 = V_1 -\left\lbrack \alpha/2 +{1\over 3}\left(\alpha_\parallel -\alpha_\perp\right)P_2(\cos\theta)\right\rbrack / R^4,
\end{eqnarray}
where $R$ is the distance between (C$_3$)-(C$_3$H) center-of-mass and the Mg atom, and $\theta$ is the angle between the vector ${\bf R}$ and the internuclear (C$_3$)-(C$_3$H) vector, ${\bf r}$. In this potentials the values of polarizabilities, $\alpha,\alpha_\parallel,\alpha_\perp$ (isotropic, parallel and perpendicular to the internuclear axis), and electric dipole,  $d$, for C$_6$H are taken from \cite{Woon-Herbst:09}.

The MDwQT code \citep{Sanz-Sanz-etal:15} is used, treating the system as three effective atoms in cartesian coordinates. A harmonic potential with large constant is added  to keep (C$_3$)-(C$_3$H) rigid. The capture rate coefficient is calculated as
\begin{eqnarray}
K_{QCT}(T) =\sqrt{ {8K_BT\over \pi\mu}} \, \pi b^2(T) \, P_c(T),
\end{eqnarray}
where $\mu =m_{Mg}m_{C_6H}/(m_{Mg}m_{C_6H})$, $b(T)$ is the maximum impact parameter and $P_c(T)=N_c/N_{tot}$ is the capture probability. $N_{tot}$=10$^5$ is the total number of trajectories per temperature and $N_c$ are those trajectories that reach a value $R<R_{capture}=$ 5 \AA. The trajectories are started at $R_0$= 600 \AA. Translational energy and initial C$_6$H  rotational angular momentum are sampled using a Monte Carlo method according to Boltzmann energy distributions. At each translational energy $E$, the initial impact parameter is sampled between 0 and $b_{max}$ (the maximum impact parameter obtained from Eq.~(\ref{eq:capture-xsection}), according to a quadratic on $b$, yielding $b(T=10K)=$ 460 \AA. No trapped trajectories are stopped when $R> 650$ \AA.

\end{appendix}

\end{document}